\documentclass[a4paper,11pt,nofootinbib,superscriptaddress]{revtex4-1}

\usepackage{braket}
\usepackage{amsmath}
\usepackage{amssymb}
\usepackage[dvipdfmx]{graphicx}

 \usepackage{graphicx}
 \usepackage{color}
 \usepackage{verbatim}
 \usepackage{wrapfig}
\usepackage{bm}
\usepackage{version}
\usepackage{here}

\numberwithin{equation}{section}

\begin{document}


\title{MeV-scale reheating temperature and thermalization of
  oscillating neutrinos by radiative and hadronic decays of
  massive particles}


\author{Takuya Hasegawa} 
\affiliation{\small Theory Center, IPNS, KEK, Tsukuba 305-0801, Japan}
\affiliation{\small The Graduate University of Advanced Studies (Sokendai), Tsukuba 305-0801, Japan}
\author{Nagisa Hiroshima}
\affiliation{\small Theory Center, IPNS, KEK, Tsukuba 305-0801, Japan}
\affiliation{\small RIKEN Interdisciplinary Theoretical and Mathematical Sciences (iTHEMS), Wako, Saitama 351-0198, Japan}
\affiliation{\small Institute for Cosmic Ray Research, University of Tokyo, Kashiwa, Chiba 277-8582, Japan}
\author{Kazunori Kohri}
\affiliation{\small Theory Center, IPNS, KEK, Tsukuba 305-0801, Japan}
\affiliation{\small The Graduate University of Advanced Studies (Sokendai), Tsukuba 305-0801, Japan}
\affiliation{ \small
  Kavli IPMU (WPI), UTIAS, The University of Tokyo, Kashiwa,
  Chiba 277-8583, Japan}
\author{Rasmus S. L. Hansen}
\affiliation{\small Max-Planck-Institut f\,$\ddot{u}$r Kernphysik, Saupfercheckweg 1, 69117 Heidelberg, Germany}
\affiliation{\small Department of Physics and Astronomy, University of Aarhus, Ny Munkegade 120, DK–8000 Aarhus C, Denmark}
\author{Thomas Tram}
\affiliation{\small Department of Physics and Astronomy, University of Aarhus, Ny Munkegade 120, DK–8000 Aarhus C, Denmark}
\affiliation{\small Aarhus Institute of Advanced Studies (AIAS), Aarhus University, DK–8000 Aarhus C, Denmark}
\author{Steen Hannestad}
\affiliation{\small Department of Physics and Astronomy, University of Aarhus, Ny Munkegade 120, DK–8000 Aarhus C, Denmark}
\date{\today}


\begin{abstract}
 From a theoretical point of view, there is a strong motivation to consider 
 an MeV-scale reheating temperature induced by long-lived massive particles 
 with masses around the weak scale, decaying only through gravitational interaction. 
 In this study, we investigate lower limits on the reheating temperature 
 imposed by big-bang nucleosynthesis assuming both radiative and hadronic 
 decays of such massive particles. For the first time, effects of neutrino 
 self-interactions and oscillations are taken into account in the neutrino 
 thermalization calculations. By requiring consistency between theoretical 
 and observational values of light element abundances, we find that the 
 reheating temperature should conservatively be $T_{\rm RH} \gtrsim 1.8$~MeV 
 in the case of the 100\% radiative decay, and $T_{\rm RH} \gtrsim$ 4--5~MeV 
 in the case of the 100\% hadronic decays for particle masses in the range 
 of 10~GeV to 100~TeV.
\end{abstract}


\maketitle

%
\section{\label{Sec:intro}Introduction} 
In the standard big-bang cosmology it is normally assumed that radiation 
components (photons, electrons/positrons, and neutrinos) were perfectly 
thermalized, and energy of radiation dominated the total energy density of 
the Universe well before the beginning of big-bang nucleosynthesis (BBN). 
In a modern picture of the early Universe, this radiation-dominated epoch 
is expected to be realized after a decay of a massive particle such as the 
inflaton, the particle associated with the inflaton field driving inflation, 
or another massive particle such as the curvaton. If such massive particles 
abundantly existed in the early Universe, their non-relativistic energy 
could dominate the total energy, and then an early matter-dominated epoch 
should have been realized before the radiation-dominated epoch. 
Therefore, particle production caused by their decays and subsequent 
entropy production (called reheating) dramatically modify the thermal 
history of the Universe. The Universe could experience the reheating 
more than once after inflation depending on the fundamental theory 
of particle physics. Since many theoretical models have been proposed 
as a theory beyond the standard model of particle physics, 
it is required to have some ways to find the true theory of nature. 
The one of the approaches is to investigate the possible value of 
``reheating temperature'' which is defined by the cosmic temperature 
when the radiation-dominated epoch just started. This is because the 
reheating temperature is related to the property of the massive particles, 
and we can constrain the theories through the observational bound on the 
reheating temperature.

As for a candidate of inflaton field or curvaton field, a lot of 
unstable massive scalar fields, {\it e.g.} moduli, dilaton fields, 
are predicted in particle physics theories beyond the standard model 
such as supergravity or superstring theory. They tend to dominate 
the total energy of the Universe during their oscillation epochs. 
It is notable that they typically have masses at or above the 
weak scale and decay only through gravitational interaction. 
This means that they have long lifetimes of ${\cal O}(0.1)$~sec--${\cal O}(10)$~sec, 
and the reheating temperature after their decay is expected to be ${\cal O}(1)$~MeV.
Since neutrinos decoupled from the thermal plasma at around the cosmic temperature 
$T \sim {\cal O}(1)$~MeV, they would have suffered imperfect thermalization 
due to the late-time entropy production caused by their decay. 
Thus, we have a strong motivation to observationally test this kind of 
cosmological scenarios with decaying particles which induce the MeV-scale 
reheating temperature.

The theory of BBN based on the standard big-bang cosmology, {\it i.e.}
standard BBN, successfully explains observational light element 
abundances (see {\it e.g.} Ref.~\cite{PDG2018} and references therein), 
and the theory say that the light elements are synthesized at around 
$T \sim {\cal O}(0.01)$~MeV--${\cal O}(0.1)$~MeV. 
As we shall see in Sec.~\ref{Sec:results_BBN}, BBN is highly sensitive 
to the neutrino abundances. Therefore, we can examine the MeV-scale 
reheating scenarios by using BBN as a probe.

In Ref.~\cite{Kawasaki_1999}, lower bounds on reheating temperature
have been studied in terms of BBN for the first time. They have 
shown that an incomplete thermalization of neutrinos gives the 
most significant effect on BBN assuming that the 100\% of the 
long-lived massive particles decay into electromagnetic radiations 
such as photons or charged leptons. Because of a competition between 
decreases and increases of the produced amount of $^4$He by the 
imperfect thermalization of the neutrinos, we can constrain the 
reheating temperature. As a result, they have obtained a conservative 
lower bound on the reheating temperature $T_{\rm RH} > 0.5$~MeV--0.7~MeV (95\% C.L.).

Afterwards, in Ref.~\cite{Kawasaki_2000}, they discussed hadronic 
decays of massive particles, {\it i.e.} direct decays into quarks 
and/or gluons which immediately fragment into hadrons such as pions, 
kaons, or nucleons. The thermalization of radiations proceed in 
exactly the same way as in the case where the 100$\%$ of the massive 
particles decay into electromagnetic radiations. This is because 
almost all of the kinetic energy of hadrons are transferred into 
radiation through Coulomb scattering with background electrons/positrons 
or inverse-Compton like scattering with background photons, and 
a neutral pion $\pi^0$ immediately decays into two photons. 
In the case of the hadronic decay, interconverting reactions 
between ambient protons and neutrons induced by emitted hadrons 
are extraordinarily important because they increase the neutron 
to proton ratio which is a key parameter of resultant abundances 
of light elements. As a result, they obtained a lower bound 
$T_{\rm RH} >$ 2.5--4~MeV (95\% C.L.) depending on the mass 
of the long-lived massive particles and their branching ratio 
into hadrons.

Subsequently, two- and three-flavor neutrino oscillations were 
respectively considered in the thermalization process of neutrinos 
in Refs.~\cite{Ichikawa_2005} and~\cite{Salas_2015} where they 
obtained a lower bound $T_{\rm RH} > 2$~MeV (95\% C.L.) and 
$T_{\rm RH} > 4.1$~MeV (95\% C.L.) assuming radiative decay 
of the massive particles. 

Some other cosmological probes other than BBN are also sensitive 
to neutrino abundances. Here we briefly refer to the recent 
papers which focused on this topic. In Ref.~\cite{Kawasaki_2000}, 
they discussed possible effects of an incomplete thermalization 
of neutrinos on a temperature anisotropy and polarization of 
Cosmic Microwave Background~(CMB) and a galaxy power spectrum 
of Large Scale Structure~(LSS). Ref.~\cite{Hannestad_2004} obtained 
a combined constraint $T_{\rm RH} > 4$~MeV (95\% C.L.) by considering 
BBN, CMB~(WMAP) and LSS~(2dF Galaxy Redshift Survey). 

After that, authors in Ref.~\cite{Ichikawa_2006} have updated the 
CMB and LSS data by using WMAP three-year data and SDSS luminous 
red galaxies data, and they obtained $T_{\rm RH} > 2$~MeV (95\% C.L.). 
Similar analysis have beendone in Ref.~\cite{Bernardis_2008} by using 
WMAP five-year data and SDSS luminous red galaxies data where they 
obtained $T_{\rm RH} > 2$~MeV (95\% C.L.) from CMB only and 
$T_{\rm RH} > 3.2$~MeV (95\% C.L.) from CMB by including external 
priors from SDSS red luminous galaxy survey and the constraint from 
the comic age. In addition, authors in Ref.~\cite{Salas_2015} 
have reported a new constraint $T_{\rm RH} > 4.3$~MeV (95\% C.L.) 
from CMB by using Planck 2015 data.

In this paper, we extend the study of Ref.~\cite{Kawasaki_2000} 
by considering neutrino oscillation and neutrino self-interaction
in the calculation of the neutrino thermalization. We assume both 
radiative and hadronic decays of the massive particles and give an 
updated bound on the reheating temperature set by BBN. This is the 
first study that consider effects of neutrino self-interactions on 
the neutrino thermalization to constrain the reheating temperature.

The structure of this paper is as follows. In Sec.~\ref{Sec:dynamics}, 
we introduce the formalism of the neutrino thermalization assuming 
the MeV-scale reheating temperature. In Sec.~\ref{Sec:results_Reheating}, 
we show the results of neutrino thermalization in the reheating, 
and describe how neutrino oscillation and neutrino self-interaction 
affect the thermalization process. The results of BBN are shown in 
Sec.~\ref{Sec:results_BBN} where we discuss effects of both radiative 
and hadronic decays on light element abundances. 
Finally, we draw our conclusion in Sec.~V.

\section{\label{Sec:dynamics}Reheating and Neutrino thermalization}
In this section, we describe the neutrino thermalization in the
low-reheating-temperature Universe and introduce the key equations.

As described in the previous section, there are some candidate
particles in theories going beyond the Standard Model of particle
physics which are weakly interacting and decay at around BBN. 
Here, we call the long-lived massive particles just ``massive particles'' 
and label them $X$. We assume the energy density of the massive particles 
dominates those of other particles at an initial time and the Universe is 
completely matter-dominated before the massive particles start to decay.  
In this setting, the entropy production caused by the decay of $X$ induces 
particle production via electromagnetic, weak, and strong interactions. 
As a result, photons and charged leptons are rapidly thermalized in the
reheating via electromagnetic interactions, whereas the weakly
interacting neutrinos are slowly created in the thermal bath of
photons and charged leptons.~\footnote{
  In the case where the massive
  particles decay intro quarks and/or gluons, they fragment into mesons 
  and baryons after the hadronization, and almost all the 
  kinetic energy of hadrons are transferred into radiation due to 
  the reason mentioned before. Therefore, contributions of secondary 
  neutrinos produced by the decay of hadrons can be neglected. 
  However, there is also another possibility that the massive particles 
  directly decays into neutrinos, {\it e.g.}
  $X \rightarrow \nu_\alpha + \bar{\nu}_\alpha$ where
  $\alpha = e, \mu, \tau$~\cite{Hannestad_2004}. In this case,
  electromagnetic plasma is produced only from neutrinos via weak
  interaction, which gives totally different results of the 
  neutrino thermalization and BBN. We do not consider the 
  possibility in this paper.
  }  
Since neutrinos decouple from the thermal bath at around a temperature 
of ${\cal O}(1)$~MeV, neutrinos should not be fully thermalized if 
$T_{\rm RH} \sim {\cal O}(1)$~MeV. A degree of
thermalization of neutrinos affects the light element
abundances~\cite{Kawasaki_1999}. For this reason, it is especially
important to accurately calculate the thermalization of neutrinos
in the reheating for $T_{\rm RH} \sim {\cal O}(1)$~MeV. Therefore,
next we look into the dynamics of neutrino thermalization 
in the thermal plasma.

In the Universe with a temperature of ${\cal O}(1)$~MeV, electrons and
positrons are the only charged leptons which are abundant in the
system since the abundances of muons and tau leptons are strongly
suppressed by Boltzmann factors. Therefore, neutrinos are mainly
produced in the annihilation process of electrons and positrons,
$e^- + e^+ \rightarrow \nu_\alpha + \bar{\nu}_\alpha$ where
$\alpha = e,\,\mu,\,\tau$. Since electron neutrino ($\nu_e$) is not
only produced by the neutral-current weak interaction but also by the
charged-current one, it tends to be produced more than $\mu$ neutrinos
($\nu_\mu$) and $\tau$ neutrinos ($\nu_\tau$) when all neutrinos are
not fully thermalized. Consequently, neutrino oscillations play a
role in equilibrating neutrino abundances in this case,
and we have to simultaneously consider the neutrino production by
collisions and flavor oscillations.

Our treatment of neutrino oscillation is consistent with that of
Ref.~\cite{Ichikawa_2005}. That is, we adopt the effective
two-flavor mixing scheme which is a good description to approximately
include full three-flavor mixings when the collision rates of $\nu_\mu$
and $\nu_\tau$ are identical, and one mixing angle is predominantly
important compared to others (see {\it e.g.} Ref.~\cite{Johns_2016} for
more details on the effective two-flavor mixing scheme). The former
condition is well satisfied because of the absence of muons and tau
leptons in the system with a temperature of ${\cal O}(1)$~MeV. On the other
hand, the latter condition is only approximately satisfied since the
reactor neutrino mixing $\theta_{13}$ is known to be non-negligible
compared to other mixings, namely the solar neutrino mixing
$\theta_{12}$ and the atmospheric neutrino mixing $\theta_{23}$~\cite{Salas_2017}. 
In later sections, however, we show that the effect 
of $\theta_{13}$ on BBN is very small compared to that of $\theta_{12}$
or neutrino self-interaction irrespective of the mass ordering of
neutrinos. Therefore, the effective two-flavor mixing scheme (with the
solar neutrino mixing) gives a good description of the full
three-flavor mixings at least for the current purpose. In this scheme,
we label the degenerate state of $\nu_\mu$ and $\nu_\tau$ as $x$ neutrinos 
($\nu_x$) and consider the flavor mixing between $\nu_e$ and
$\nu_x$. Also, we label the other neutrino species which does not mix
with other flavors as spectator neutrino ($\nu_{\rm sp}$) in this
two-flavor treatment.

In general, neutrino states can be described by a one-particle
irreducible density matrix $\varrho_{\bm p} \equiv \varrho(p,t)$ 
where $p \equiv |{\bm p}|$ is the absolute momentum.~\footnote{
  Since we focus on the Universe with a temperature of ${\cal O}(1)$~MeV, 
  we can neglect tiny neutrino masses which are known to be 
  sub-eV scale~\cite{PDG2018}. In this case, energy of neutrinos 
  $E$ is equal to its absolute momentum, {\it i.e.} $E = p$.
  }  
Since we focus on the effective two-flavor mixing, 
the density matrix is expressed in terms of a $2 \times 2$ Hermitian matrix, 
and we label each element of the density matrix as 
\begin{eqnarray}
\varrho_{\bm p} = \left(
\begin{array}{cc}
\rho_{ee} & \rho_{ex}\\
\rho^*_{ex} & \rho_{xx}
\end{array}
\right)\,. 
\end{eqnarray}
The diagonal part of the matrix corresponds to the distribution
function of mixed neutrinos ({\it i.e.}~$\nu_e$ and $\nu_x$), that is,
$\rho_{ee} = f_{\nu_e}$ and $\rho_{xx} = f_{\nu_x}$, while the
off-diagonal elements represent the quantum coherence among neutrinos
with different flavors. In the current study, the chemical potentials
of neutrinos are set to be zero. Under this assumption, the density
matrix of neutrinos is equal to that of anti-neutrinos, {\it i.e.}
$\varrho_{\bm p} = \bar{\varrho}_{\bm p}$, and they have 
the same abundance. Therefore, it is not necessary to follow the 
time evolution of anti-neutrinos separately from that of the 
corresponding neutrinos.

We can obtain the time evolution of the neutrino density matrix by
solving the momentum-dependent Quantum Kinetic Equation~\cite{McKellar_1994,Sigl_1993} 
which is formally written as 
\begin{equation}
 \frac{d\varrho_{\bm p}}{dt} = \frac{\partial\varrho_{\bm p}}{\partial t} - H\, p 
  \frac{\partial\varrho_{\bm p}}{\partial p} = -i\,[{\mathcal H}_{\bm p},\varrho_{\bm p}] 
  + C(\varrho_{\bm p})\,.
\label{Eq:QKE}
\end{equation}
In the above equation, the term including the Hubble parameter $H$
corresponds to the effect of the expansion of the Universe, and
$C(\varrho_{\bm p})$ is the collision term of neutrinos expressed as 
\begin{eqnarray}
C(\varrho_{\bm p}) = \left(
\begin{array}{cc}
R_{\nu_e} & -D \rho_{ex}\\
-D \rho^*_{ex} & R_{\nu_x}
\end{array}
\right)\,, 
\end{eqnarray}
where $R_{\nu_e}$ and $R_{\nu_x}$ are the production rates of $\nu_e$
and $\nu_x$, respectively. Also, $D$ is the collisional-damping rate
which breaks the flavor coherence among different flavors of
neutrinos. In this paper, we adopt a simplified treatment of the
damping effects discussed in Ref.~\cite{Salas_2015} and neglect the
additional contributions such as ``damping-like terms'' which appear
in Ref.~\cite{McKellar_1994}.  In the current study, we consider the
collisional processes $a(k) + b(p) \rightarrow c(k') + d(p')$
shown in Table.~I including those of neutrino self-interaction.  
In this case, the expressions of the repopulation and the damping terms
are~\cite{Hannestad_2015},
\begin{equation}
 \begin{aligned}
  R_{\nu_\alpha}(k) &=  2\pi \int d \Pi_{k'} d \Pi_{p'}d \Pi_{p} \; \delta_E(kp| k'p') \\
  &\hspace{1.5cm} \times \sum_i \mathcal{V}^2[\nu_\alpha(k),\bar \nu_\alpha (p)| i(k'),\bar i(p')] 
  \left[f_i(E_{k'})f_{\bar{i}}(E_{p'})(1-f_{\nu_\alpha}(k))(1- f_{\bar\nu_{\alpha}}(p))\right. \\
  &\hspace{5cm} - \left.f_{\nu_\alpha}(k) f_{\bar\nu_{\alpha}}(p)(1-f_i(E_{k'}))(1-f_{\bar{i}}(E_{p'}))\right] \\
  &\hspace{1.9cm} + \sum_j\mathcal{V}^2[\nu_\alpha(k),j (p)| \nu_\alpha(k'), j(p')] 
  \left[f_{\nu_\alpha}(k')f_j(E_{p'})(1-f_{\nu_\alpha}(k)) (1-f_j(E_p))\right.\\
  &\hspace{5cm}\left.- f_{\nu_\alpha}(k) f_j(E_p)(1-f_{\nu_\alpha}(k'))(1-f_j(E_{p'}))\right] \,,
  \label{Eq:Ralpha}
 \end{aligned}
\end{equation}
\begin{equation}
 \begin{aligned}
  D(k) &= \pi\, \sum_{\alpha} \int d \Pi_{k'} d \Pi_{p'} d \Pi_{p} \; \delta_E(kp|k'p') \\
  &\hspace{1.5cm} \times \sum_i \mathcal{V}^2[\nu_\alpha(k),\bar\nu_\alpha(p)| i(k'),\bar i(p')] 
  \left[f_i(E_{k'})f_{\bar i}(E_{p'}) (1-f_{\bar \nu_\alpha}(p)) \right. \\
  &\hspace{5cm} \left.+ f_{\bar \nu_\alpha}(p) (1-f_{\bar i}(E_{p'})) (1-f_{i}(E_{k'})) \right] \\
  &\hspace{1.9cm} + \sum_j \mathcal{V}^2[\nu_\alpha(k),j(p)|\nu_\alpha(k'),j(p')] 
  \left[f_{\nu_\alpha}(k')f_j(E_{p'}) (1-f_j(E_p))\right.  \\
  &\hspace{5cm} \left.+ f_j(E_p) (1-f_j(E_{p'})) (1-f_{\nu_\alpha}(k'))\right] \,, 
  \label{Eq:Deq}
 \end{aligned}
\end{equation}
where $\alpha = e, x$, and $k, p, k'$ and $p'$ are absolute momenta 
of the particle $a, b, c$ and $d$, respectively. 
Also, $d \Pi_p \equiv \frac{d^3{\bm p}}{(2\pi)^3}$, and
$\delta_E(k p | k' p') \equiv \delta^{(1)} (E_k+E_p-E_{k'}-E_{p'})$ is
the 1D Dirac delta function corresponding to energy conservation for
each process. The summation index $i$ runs over electrons and all
flavors of neutrinos other than $\nu_\alpha$ ({\it i.e.} $\nu_\beta$ where
$\beta \neq \alpha$), while $j$ runs in addition over positrons,
$\nu_\alpha$, and all flavors of anti-neutrinos. The expression of
$\mathcal{V}^2$ is written as 
\begin{equation}
 \mathcal{V}^2[a(p),b(k)| c(p'),d(k')] = (2\pi)^3 \delta^{(3)}(k+p,k'+p') 
  N^2_aN^2_bN^2_cN^2_d S|M|^2(a(p),b(k)| c(p'),d(k'))\,,
\end{equation}
where $S|M|^2(a(p),b(k)| c(p'),d(k'))$ is the squared scattering matrix element 
for the processes in Table.~I summed over initial and final spins, 
and symmetrized over identical particles in the initial and the final state.
Also, $N_i \equiv \sqrt{1/2E_i}$ where $E_i$ is the energy of particle $i$\, 
($i = a,\,b,\,c,\,d$) and $\delta^{(3)}(k+p,k'+p') \equiv \delta^{(3)}({\bm k} 
+ {\bm p} - {\bm k'} - {\bm p'})$ is the 3D Dirac delta function corresponding 
to the momentum conservation. As for the processes in Table.~I, 
we analytically reduce the dimension of momentum integrals in the above 
expressions from nine to two and calculate the full collision terms without 
any simplifying assumptions in the same way as in Ref.~\cite{Hannestad_2015}.

\begin{table*}
  \begin{tabular}{cc}
    \hline
    \hline
    Process ($\alpha \neq \beta$) & $ S\, |M|^2$ \\
    \hline
    \raisebox{5.5pt}{I.$\ \ \ \ e^-+e^+ \rightarrow \nu_\alpha+\bar{\nu}_\alpha$}
    & \shortstack{\ \ $2^5\, G_{\rm F}^{2}\, [\,(2\sin^2\theta_W \pm 1)^2(k\cdot p')(p\cdot k') + 4\sin^4\theta_W (k\cdot k') (p\cdot p')\ \ $\\
    $\ \ \ \ \ \ \ \ \ \ \ \ \ \ \ \ \ \ \ \ \ \ \ \ \ \ \ \ \ \ \ \ \ +2\,m_e^2\sin^2\theta_W (2\sin^2\theta_W \pm 1) (k\cdot p)\,]$}\\
    \hline
    \raisebox{5.5pt}{I\hspace{-.1em}I.$\ \ \ \ \nu_\alpha + e^+ \rightarrow \nu_\alpha + e^+$}
    & \shortstack{\ \ $2^5\, G_{\rm F}^{2}\, [\,(2\sin^2\theta_W \pm 1)^2 (k\cdot k') (p\cdot p') + 4\sin^4\theta_W (k\cdot p) (k'\cdot p')\ \ $\\
    $\ \ \ \ \ \ \ \ \ \ \ \ \ \ \ \ \ \ \ \ \ \ \ \ \ \ \ \ \ \ \ \ \ -2\,m_e^2\sin^2\theta_W (2\sin^2\theta_W \pm 1) (k\cdot p')\,]$}\\ 
    \raisebox{5.5pt}{I\hspace{-.1em}I\hspace{-.1em}I.$\ \ \ \ \nu_\alpha + e^- \rightarrow \nu_\alpha + e^-$}
    & \shortstack{\ \ $2^5\, G_{\rm F}^{2}\, [\,(2\sin^2\theta_W \pm 1)^2 (k\cdot p) (k'\cdot p') + 4\sin^4\theta_W (k\cdot k') (p\cdot p')\ \ $\\
    $\ \ \ \ \ \ \ \ \ \ \ \ \ \ \ \ \ \ \ \ \ \ \ \ \ \ \ \ \ \ \ \ \ -2\,m_e^2\sin^2\theta_W (2\sin^2\theta_W \pm 1) (k\cdot p')\,]$}\\
    \hline
    I\hspace{-.1em}V.$\ \ \ \ \nu_\alpha+\nu_\alpha\rightarrow\nu_\alpha+\nu_\alpha$
    & $2^6\, G_{\rm F}^{2}\, (k\cdot p) (k'\cdot p')$\\ 
    V.$\ \ \ \ \nu_\alpha+\nu_\beta\rightarrow\nu_\alpha+\nu_\beta$
    & $2^5\, G_{\rm F}^{2}\, (k\cdot p) (k'\cdot p')$\\ 
    V\hspace{-.1em}I.$\ \ \ \ \nu_\alpha+\overline{\nu}_\alpha\rightarrow\nu_\alpha+\overline{\nu}_\alpha$
    & $2^7\, G_{\rm F}^{2}\, (k\cdot p') (p\cdot k')$\\ 
    V\hspace{-.1em}I\hspace{-.1em}I.$\ \ \ \ \nu_\alpha+\overline{\nu}_\beta\rightarrow\nu_\alpha+\overline{\nu}_\beta$
    & $2^5\, G_{\rm F}^{2}\, (k\cdot p') (p\cdot k')$\\ 
    V\hspace{-.1em}I\hspace{-.1em}I\hspace{-.1em}I.$\ \ \ \ \nu_\alpha+\overline{\nu}_\alpha\rightarrow\nu_\beta+\overline{\nu}_\beta$
    & $2^5\, G_{\rm F}^{2}\, (k\cdot p') (p\cdot k')$\\ \hline\hline
  \end{tabular}
 \caption{
 \label{nu-process}
 Collision process $a(p) + b(k) \rightarrow c(p') + d(k')$ which contributes to 
 the thermalization of neutrinos of each flavor $\nu_\alpha, \nu_\beta$ 
 ($\alpha,\beta = e, \mu, \tau\ {\rm where}\ \alpha \neq \beta$). 
 The process I is the production process of neutrinos due to electron annihilation, 
 the processes I\hspace{-.1em}I--I\hspace{-.1em}I\hspace{-.1em}I are the scattering 
 processes between neutrinos and electrons, and the processes 
 I\hspace{-.1em}V--V\hspace{-.1em}I\hspace{-.1em}I\hspace{-.1em}I are 
 the self-interaction processes among neutrinos. Here, $\theta_W$ is the 
 Weinberg angle, $G_{\rm F}$ is the Fermi-coupling constant, $S$ is the symmetry factor, 
 and $|M|^2$ is the squared scattering matrix element. The positive sign in the expression 
 is for $\nu_e$ and the minus sign for $\nu_\mu$ or $\nu_\tau$ ({\it i.e.} for $\nu_x$ 
 and $\nu_{\rm sp}$). The expressions of the process I, I\hspace{-.1em}I, I\hspace{-.1em}V 
 and V are also applied to the corresponding anti-neutrinos.
 }
\end{table*}

In the expression of Eq.~\eqref{Eq:QKE}, ${\mathcal H}_{\bm p}$ is neutrino Hamiltonian 
which is expressed as 
\begin{equation}
 {\mathcal H}_{\bm p} = {\mathcal H}_{{\bm p},\,{\rm vac}} 
  + {\mathcal H}_{{\bm p},\,{\rm mat}} = \frac{{\sf M}^2}{2p} -\frac{8\sqrt{2}\,G_{\rm F}p}{3} 
  \left[ \frac{\bm{E}_l}{m^2_W} + \frac{\bm{E}_\nu}{m^2_Z} \right] 
  + \sqrt{2}\,G_{\rm F}\int d \Pi_{p'} (\varrho_{\bm p'} - \bar{\varrho}_{\bm p'}^*)\,,
\label{Eq:hamiltonian}
\end{equation}
where $G_{\rm F}$ is the Fermi-coupling constant. Also, $m_W$ and $m_Z$ are the masses 
of $W$ and $Z$ bosons, respectively. In the above expression, the first term, 
${\mathcal H}_{{\bm p},\,{\rm vac}}$, is the contribution which induces the vacuum oscillation 
where ${\sf M}$ is the mass matrix in flavor basis. The mass matrix ${\sf M}$ is related 
to the one in mass basis $\mathcal M$ as ${\sf M^2} = U {\mathcal M^2} U^\dagger$ where 
$U$ is the PMNS matrix. In the effective two-flavor mixing scheme, 
\begin{equation}
{\mathcal M^2} = \left(
\begin{array}{cc}
m_1^2 & 0 \\
0 & m_2^2
\end{array} \right)\ ,\ 
  U = \left(
\begin{array}{cc}
\cos\theta & \sin\theta \\
-\sin\theta & \cos\theta
\end{array} \right) \,,
\end{equation}
where $\delta m^2 \equiv m_2^2 - m_1^2$ is the squared-mass difference and 
$\theta$ is the mixing angle in vacuum between $\nu_e$ and $\nu_x$. Also, 
the second term in the Hamiltonian, ${\mathcal H}_{{\bm p},\,{\rm mat}}$, 
corresponds to the matter potentials which arise from coherent scatterings 
between neutrinos and charged-leptons. In the term, ${\bm{E}_l}$ corresponds 
to the total energy density of charged leptons, while ${\bm{E}_\nu}$ to that 
of neutrinos:
\begin{equation}
{\bm{E}_l} = \left(
\begin{array}{cc}
\rho_e & 0 \\
0 & 0
\end{array} \right)\ ,\
{\bm{E}_\nu} = \int\!d \Pi_{p'}\, p'\,(\varrho_{\bm p'} + \bar{\varrho}_{\bm p'}^*) = \left(
\begin{array}{cc}
\rho_{\nu_e} & \rho_{\nu_{ex}} \\
\rho^*_{\nu_{ex}} & \rho_{\nu_x}
\end{array} \right) \,,
\label{Eq:matterpote}
\end{equation}
where $\rho_e = \rho_{e^-} + \rho_{e^+}$ 
is the total energy density of electrons and positrons, while $\rho_{\nu_e}$ 
and $\rho_{\nu_x}$ are those of $\nu_e$ and $\nu_x$, respectively. Also, we have defined 
$\rho_{\nu_{ex}} \equiv \int\! d \Pi_{p'}\, p'\, (\rho_{ex} + \bar{\rho}_{ex}^*)$ 
and $\rho^*_{\nu_{ex}} \equiv \int\! d \Pi_{p'}\, p'\,(\rho^*_{ex} + \bar{\rho}_{ex})$. 
In the expression of ${\bm{E}_l}$, we have neglected the existence of muons or tau leptons 
due to their large masses.~\footnote{
  In the effective two-flavor mixing scheme, 
  we need to treat both $\nu_\mu$ and $\nu_\tau$ in the same way. Therefore, 
  we do not consider background muons or tau leptons whose contribution 
  is very small compared to that of electrons.
  }
The asymmetric part of ${\mathcal H}_{{\bm p},\,{\rm mat}}$ is often assumed 
to vanish when $\rho = \bar{\rho}$ for neutrinos and the number density of 
electrons and positrons are identical. For the diagonal part of the Hamiltonian 
this is true, however the off-diagonal part gets a contribution from the 
neutrinos as shown in Eq.~\eqref{Eq:hamiltonian} since $\varrho_{\bm p}^* \neq \varrho_{\bm p}$.

As for the oscillation parameters, we use the best fit values of the 
mass-squared differences and mixing angles reported in 
Ref.~\cite{Salas_2017}:~\footnote{
  The atmospheric neutrino mixing $(\theta_{23}, 
  \delta m_{23}^2)$ is irrelevant to the oscillation between $\nu_e$ and 
  $\nu_x$ in the effective two-flavor mixing scheme. Therefore, we do not 
  use the value in this paper.
}
\begin{eqnarray}
 \delta m _ { 12 } ^ { 2 }  &=& 7.55 \times 10 ^ { - 5 }~{\rm eV} ^ { 2 }\ ,
  \ \ \sin ^ { 2 } \theta _ { 12 } = 3.20 \times 10^{-1} \,, \label{mixing_para1}\\
 \delta m _ { 13 } ^ { 2 }  &=& 2.50 \times 10 ^ { - 3 }~{\rm eV } ^ { 2 }\ ,
  \ \ \sin ^ { 2 } \theta _ { 13 } = 2.160 \times 10^{-2}\ \ {\rm (NO)}\,,\label{mixing_para2}\\
 \delta m _ { 13 } ^ { 2 }  &=& - 2.42 \times 10 ^ { - 3 }~{\rm eV } ^ { 2 }\ ,
  \ \ \sin ^ { 2 } \theta _ { 13 } = 2.220 \times 10^{-2}\ \ {\rm (IO)}\ \label{mixing_para3},
\end{eqnarray}
where ``NO'' (``IO'') means normal (inverted) mass ordering of neutrinos, respectively.
For the numerical calculation, we rewrite the $2\times2$ density matrix with 
polarization vectors ($P_0,\boldsymbol{P}$):
\begin{equation}
 \varrho_{\bm p} = \left(
\begin{array}{cc}
 \rho_{ee} & \rho_{ex} \\
 \rho^*_{ex} & \rho_{xx}
\end{array} \right)
 = \frac{1}{2}\,[\, P_0(p)\,\sigma_0 + \boldsymbol{P}(p)\,\cdot {\bm \sigma}\,] \,,
\end{equation}
where $\sigma_0 = \bm{1}$ is the identity matrix and 
${\bm \sigma} = (\sigma_x, \sigma_y, \sigma_z)$ are the Pauli matrices. 
With $\boldsymbol{P} = (P_x,P_y,P_z)$, distribution functions of mixed 
neutrinos can be written as 
\begin{equation}
 f_{\nu_e} = \frac{1}{2}(P_0 + P_z)\ ,\ f_{\nu_x} = \frac{1}{2}(P_0 - P_z) \,.
\end{equation}
In addition, we can rewrite the expression of Eq.~\eqref{Eq:QKE} as follows:
\begin{eqnarray}
 \dot{\boldsymbol{P}} &=& \overrightarrow{\mathcal{H}} \times \boldsymbol{P} 
 - D \,(P_x\, \mathbf{x}+P_y\, \mathbf{y}) + (R_{\nu_e} - R_{\nu_x})\,\mathbf{z} \,,
 \label{Eq:QKE_pola}\\
 \dot{P}_0 &=& R_{\nu_e} + R_{\nu_x} \,,
  \label{Eq:QKE_pola2}
\end{eqnarray}
which leads to 
\begin{eqnarray}
 \dot{P}_{\nu_e} &=& \mathcal{H}_x\, P_y - \mathcal{H}_y\, P_x + 2\,R_{\nu_e} \,, \label{Eq:QKE_pola_P_nu_e}\\
 \dot{P}_{\nu_x} &=& \mathcal{H}_y\, P_x - \mathcal{H}_x\, P_y + 2\,R_{\nu_x} \,, \label{Eq:QKE_pola_P_nu_x}\\
 \dot{P_x} &=&  \frac{1}{2}\,\mathcal{H}_y\, (P_{\nu_e} - P_{\nu_x}) - \mathcal{H}_z\, P_y - D \,P_x\,, \label{Eq:QKE_pola_Px}\\
 \dot{P_y} &=&  \mathcal{H}_z\, P_x - \frac{1}{2}\,\mathcal{H}_x\, (P_{\nu_e} - P_{\nu_x}) - D \,P_y\,, \label{Eq:QKE_pola_Py}
\end{eqnarray}
where $\mathbf{x}, \mathbf{y}$ and $\mathbf{z}$ are unit vectors, and we defined
$P_{\nu_e} \equiv P_0 + P_z$ and $P_{\nu_x} \equiv P_0 - P_z$. Each component 
of the neutrino potential $\overrightarrow{\mathcal{H}},\, {\rm {\it i.e.}}\  
\mathcal{H}_i = {\rm Tr}\,({\mathcal H}_{\bm p}\,{\sigma_i}$) 
where $i = x, y, z$, is written as 
\begin{eqnarray}
 \mathcal{H}_x &=&  \frac{\delta m^2}{2 p} \sin 2 \theta - \frac{16\sqrt{2}\,G_{\rm F}\, p}{3\,m_Z^2}\int\! d \Pi_{p'}\, p' P_x\,,\\
 \mathcal{H}_y &=&  2 \sqrt{2}\,G_{\rm F} \int\! d \Pi_{p'}\, P_y \,,\\
 \mathcal{H}_z &=& - \frac{\delta m^2}{2 p} \cos 2 \theta + \mathcal{H}_{\rm mat} \,.
\end{eqnarray}
The second term in $\mathcal{H}_z$ is the matter contribution which is 
explicitly written as
\begin{eqnarray}
 \mathcal{H}_{\rm mat} &=& - \frac{8\sqrt{2}}{3} G_{\rm F}\, p\left[\frac{\rho_e}{m_W^2} + \frac{\rho_{\nu_e} - \rho_{\nu_x}}{m_Z^2} \right]\,,\\
 &=& - \frac{4\sqrt{2}}{3\,\pi^2} G_{\rm F}\, p \left[\frac{g_e}{m_W^2} \int_0^\infty dp'\, p'^2 \frac{E_e}{\exp(E_e/T_\gamma)+1} \right. \nonumber\\ 
  && \left. \ \ \ \ \ \ \ \ \ \ \ \ \ \ \ \ \ \ \ \ \ \ \ \ \ \ \ \ \ + \frac{g_\nu}{m_Z^2}\int^{\infty}_{0} dp'\,p'^3 (f_{\nu_e} - f_{\nu_x}) \right]\,,
\end{eqnarray}
where $T_\gamma$ is the photon temperature, and $E_e = \sqrt{p^2 + m_e^2}$ is 
the energy of electrons. Also, $g_e = 4$ is statistical degree of freedom of 
electrons and $g_\nu = 2$ is that of neutrinos of each flavor. 

With the matter potential $\mathcal{H}_{\rm mat}$, the mass-squared difference 
and the mixing angle in vacuum are modified in medium by MSW effect 
as follows~\cite{Wolfenstein_1978,Mikheyev_1989}:
\begin{eqnarray}
 \frac{\delta m^2_M}{2p} = \sqrt{\left(\frac{\delta m^2}{2p}\right)^2 \sin^2 2\theta 
  + \left(- \frac{\delta m^2}{2 p} \cos 2 \theta + \mathcal{H}_{\rm mat}\right)^2}\ \ ,\label{Eq:MSW_mass}\\
 \sin^2 2\theta_M = \frac{\left(\frac{\delta m^2}{2p}\right)^2 \sin^2 2\theta}
  {\left(\frac{\delta m^2}{2p}\right)^2 \sin^2 2\theta + \left( - \frac{\delta m^2}{2 p} \cos 2 \theta 
  + \mathcal{H}_{\rm mat}\right)^2} \ \ ,
\label{Eq:MSW_mixing}
\end{eqnarray}
where $\delta m^2_M$ and $\theta_M$ are the in-medium mass-squared difference and 
the mixing angle, respectively. We note here that in Eqs.~\eqref{Eq:MSW_mass}--\eqref{Eq:MSW_mixing} 
we simplify the expressions by neglecting the small contributions of the off-diagonal 
part of the matter potential ${\mathcal H}_{{\bm p},\,{\rm mat}}$ in Eq.~\eqref{Eq:hamiltonian} 
(see Ref.~\cite{Johns_2016} for the exact expressions of the MSW effect). 
As can be seen from Eqs.~\eqref{Eq:MSW_mass} and \eqref{Eq:MSW_mixing}, the matter potential 
$|\mathcal{H}_{\rm mat}| \propto T_\gamma^5$ dominates the vacuum one 
$|\mathcal{H}_{\rm vac}| \equiv |- \frac{\delta m^2}{2 p} \cos 2 \theta| \propto T_\gamma^{-1}$, 
{\it i.e.} $|\mathcal{H}_{\rm mat}| >> |\mathcal{H}_{\rm vac}|$, at high temperature such as 
$T_\gamma > {\cal O}(10)$~MeV and $\theta_M \sim 0$ holds for most energy modes.~\footnote{
  We assume here that the neutrinos are thermalized with photons and have a 
  temperature $T_\gamma$ for simplicity.
  } 
On the other hand, the opposite hierarchy, {\it i.e.} $|\mathcal{H}_{\rm vac}| >> |\mathcal{H}_{\rm mat}|$, 
holds at low temperature and the mixing parameters take the same values as those in vacuum: 
$\theta_M \sim \theta,\,\delta m^2_M \sim \delta m^2$. For neutrinos with momentum 
$p\, = \langle p \rangle \sim 3.15\, T_\gamma$ where $\langle \cdot \rangle$ means 
a thermal average, the level crossing between these potentials, 
{\it i.e.} $|\mathcal{H}_{\rm mat}| \sim |\mathcal{H}_{\rm vac}|$, 
occurs at the temperature $T_{\rm c}$ 
\begin{eqnarray}
 T_{\rm c}\, \sim\, G_{\rm F}^{-1/3} (\delta m^2 \cos 2\theta)^{1/6} 
  \,   \sim \, \left\{ 
\begin{array}{c}
 3~{\rm MeV} \left(\frac{\delta m_{12}^2}{2.5 \times 10^{-3}~{\rm eV}^2}\right)^{1/6} \\
 5~{\rm MeV} \left(\frac{\delta m_{13}^2}{7.5 \times 10^{-5}~{\rm eV}^2}\right)^{1/6} 
\end{array}
 \right.\,,
\end{eqnarray}
where, in the above evaluation, we have replaced $p$ with $\langle p \rangle \sim 3.15\, T_\gamma$ 
and approximated $(\cos 2\theta)^{1/6} \sim 1$ which is well satisfied for $\theta_{12}$ 
and $\theta_{13}$. Therefore, neutrino oscillation becomes effective at around a temperature 
of ${\cal O}(1)$~MeV for the solar neutrino mixing ($\theta_{12},\delta m_{12}$) and the reactor 
neutrino mixing ($\theta_{13},\delta m_{13}$), which is the reason for taking its effect 
on the neutrino thermalization into account. 

On the other hand, since $\nu_{\rm sp}$ decouple from flavor mixings of neutrinos, 
the time evolution of this neutrino species is just given by the classical Boltzmann equation:
\begin{equation}
 \frac{d f_{\nu_{\rm sp}}}{d t} = \frac{\partial f_{\nu_{\rm sp}}}{\partial t} 
  - H\, p \frac{\partial f_{\nu_{\rm sp}}}{\partial p} = C(f_{\nu_{\rm sp}}) \,, 
\label{Eq:spBoltzmann}
\end{equation}
where $f_{\nu_{\rm sp}}$ is the distribution function of $\nu_{\rm sp}$, 
and $C(f_{\nu_{\rm sp}})$ is the collision term whose expression is equal to 
that of $\nu_x$, {\it i.e.} $C(f_{\nu_{\rm sp}}) = R_{\nu_x}$ 
(see Eq.~\eqref{Eq:Ralpha}).

In order to calculate the thermalization process of neutrinos in the expanding Universe, 
we also need to compute the energy conservation equation: 
\begin{equation}
\frac { d \rho } { d t } = - 3 H ( \rho + P ) \,,
\end{equation}
which can be expressed as the time evolution of the photon temperature $T_\gamma$\,:
\begin{equation}
 \frac{dT_\gamma}{dt} = 
  - \frac{-\Gamma_X\rho_X+4H(\rho_\gamma+\rho_\nu) + 3H(\rho_e+P_e) + 
  \frac{d\rho_\nu}{dt}}{\frac{\partial \rho_\gamma}{\partial T_\gamma} |_{a(t)} 
  + \frac{\partial \rho_e}{\partial T_\gamma} |_{a(t)}} \,,
\label{Eq:ene_conserv}
\end{equation}
where $a(t)$ is the scale factor at the cosmic time $t$, $\Gamma_X$ is the 
decay rate of the massive particles, whereas $\rho$ and $P$ are the total 
energy density and the total pressure, respectively: 
\begin{eqnarray} 
 \rho &=& \rho_\gamma + \rho_e + \rho_\nu + \rho_X \,, \nonumber \\
 &=& \frac{\pi^2}{15}T_\gamma^4 + \frac{g_e}{2\pi^2} \int_0^\infty dp' \,p'^2 
  \frac{E_e}{\exp(E_e/T_\gamma)+1} \nonumber \\ && \ \ \ \ \ \ \ \ \ \ + 
  \frac{g_\nu}{2\pi^2}\int^{\infty}_{0} dp'\,p'^3 (f_{\nu_e} + f_{\nu_x} + f_{\nu_{\rm sp}}) + \rho_X\,,\label{Eq:rhotot}\\ \vspace{10pt}
 P &=& P_\gamma + P_e + P_\nu \,, \nonumber \\
 &=& \frac{\pi^2}{45}T_\gamma^4 + \frac{g_e}{6\pi^2} \int_0^\infty dp' \, 
  \frac{p'^4}{E_e}\frac{1}{\exp(E_e/T_\gamma)+1} \nonumber \\ && \ \ \ \ \ \ \ \ \ \ 
  + \frac{g_\nu}{6\pi^2}\int^{\infty}_{0} dp'\,p'^3 
  (f_{\nu_e} + f_{\nu_x} + f_{\nu_{\rm sp}}) \,.
\end{eqnarray}
Here, $\rho_\gamma (P_\gamma)$, $\rho_e (P_e)$, $\rho_\nu (P_\nu)$ and $\rho_X$ mean 
the energy density (pressure) of photons, electrons, neutrinos and the massive particles, 
respectively. The total energy density and the total pressure of neutrinos are a sum of 
three contributions: $\rho_\nu = \rho_{\nu_e} + \rho_{\nu_x} + \rho_{\nu_{\rm sp}}$,
$P_\nu = P_{\nu_e} + P_{\nu_x} + P_{\nu_{\rm sp}}$.
In addition, the Hubble parameter $H$ is obtained by solving the Friedmann equation:
\begin{equation}
 H \equiv \frac{\dot{a}}{a} = \sqrt{\frac{8\pi G\rho}{3}} \,.
\label{Eq:Friedmann}
\end{equation}
In the above expression, we can obtain the time evolution of $\rho_X$ by solving 
the Boltzmann equation of the massive particles~$X$:
\begin{equation}
 \frac{d\rho_X}{dt} = -\Gamma_X \rho_X - 3H \rho_X \,,
\end{equation}
which can be integrated analytically for non-relativistic particles~$X$:
\begin{equation}
 \frac{\rho_X}{s} = \frac{\rho_{X,0}}{s_0}\ e^{-\Gamma_X t}\,,
\label{Eq:XBoltzmann}
\end{equation}
where $\rho_{X,0}$ and $s_0$ are respectively the initial energy- and entropy density of $X$, 
and $\rho_{X,0}$ is assumed to be much larger than those of other particles, {\it i.e.}
$\rho_{X,0} >> (\rho_\gamma + \rho_e + \rho_\nu)_{t=0}$. 
In addition, $\Gamma_X$ is related to $T_{\rm RH}$ through the 
Hubble parameter $H = H(T_{\rm RH})$ as follows:
\begin{equation}
 \Gamma_X = 3 H\,.
\label{Eq:trh_def}
\end{equation}
%
Since the energy density of the Universe is dominated by radiation components after most of 
the massive particles decayed and $T_\gamma \sim T_{\rm RH}$ is realized, we can approximately 
write the Hubble parameter as
\begin{equation}
 H = \sqrt{\frac{g^* \pi^2}{90}} \frac{T_{\rm RH}^2}{m_{\rm pl}} \,,
\end{equation}
where $m_{\rm pl} \sim 2.435 \times 10^{18}$ GeV is the reduced Planck mass, 
and $g^* = 10.75$ is the relativistic degrees of freedom in the Universe with a 
temperature of ${\cal O}(1)$~MeV. Hence, the relation between $T_{\rm RH}$ 
and the decay rate of $X$ is approximately written as
\begin{equation}
T_{\rm RH} \sim 0.7 \,\left(\frac{\Gamma_X}{\rm sec^{-1}}\right)^{1/2}~{\rm MeV} \,.
\label{Eq:trh_rela} 
\end{equation}
From the above expression, we can see that $T_{\rm RH} \sim {\cal O}(1)$~MeV 
corresponds to the lifetime of the massive particles
$\tau_X = \Gamma_X^{-1} \sim {\cal O}(1)$ sec.~\footnote{
  Since the actual value of $g^*$ depends on the value of $T_{\rm RH}$,
  Eq.~\eqref{Eq:trh_def} just gives a rough estimate of when 
  the radiation-dominated epoch is realized.
  }
In order to obtain the neutrino distribution functions and a degree of the 
neutrino thermalization in the reheating, we simultaneously solve 
the Eqs.~\eqref{Eq:QKE_pola_P_nu_e}, \eqref{Eq:QKE_pola_P_nu_x}, 
\eqref{Eq:QKE_pola_Px}, \eqref{Eq:QKE_pola_Py}, \eqref{Eq:spBoltzmann}, 
\eqref{Eq:ene_conserv}, \eqref{Eq:Friedmann} and \eqref{Eq:XBoltzmann} 
from the initial time $t = 10^{-4}$~sec to the final time $t = 10^{7}$~sec 
corresponding to the cosmic time well before and after BBN, respectively. 
We find that the final results are independent of the choice of the initial time 
as long as the initial temperature of electromagnetic particles is much higher 
than $T_{\rm RH}$. To calculate neutrino thermalization processes, 
we use a modified version of LASAGNA code~\cite{Hannestad_2012,Hannestad_2013} 
which is, in the original version, a solver of ordinary differential equations 
for calculating sterile neutrino production in the early Universe. 

In the next section, we show our numerical results of the neutrino thermalization 
and BBN in the low-reheating-temperature Universe.

\section{\label{Sec:results_Reheating}Numerical results: neutrino thermalization in the reheating}
In this section, we show our numerical results of neutrino thermalization.
In order to express the time evolution of the neutrino thermalization, 
we define the effective number of neutrino species $N_{\rm eff}$:
\begin{equation}
 N_{\rm eff} = N_{{\rm eff},\, \nu_e} + N_{{\rm eff},\, \nu_x} + N_{{\rm eff},\, \nu_{\rm sp}} 
  = \sum_{\alpha\, =\, e,\,x,\,{\rm sp}} \rho_{\nu_\alpha}/\rho_{\nu_{\alpha,\, {\rm std}}} \,,
 \label{Eq:neff_def}
\end{equation}
where $N_{{\rm eff},\,\nu_\alpha}$ is the contribution for each neutrino species, 
and $\rho_{\nu_{\alpha,\, {\rm std}}}$ is the energy density of each neutrino 
species in the standard big-bang cosmology.~\footnote{
  The energy density of $\nu_e$ is slightly larger than those of 
  $\nu_x$ and $\nu_{\rm sp}$ after electron annihilation due to 
  the larger reaction rate of $\nu_e$ with electrons. Therefore, 
  we discriminate among $\rho_{\nu_{\alpha,\, {\rm std}}}$ with 
  different flavors.
  } 
The value of $N_{\rm eff}$ is almost equal to the actual number of neutrino species 
when all neutrinos are fully thermalized. 

\begin{figure}[tbph]
\begin{center}
\resizebox{23cm}{!}{\hspace{2.6cm}\input{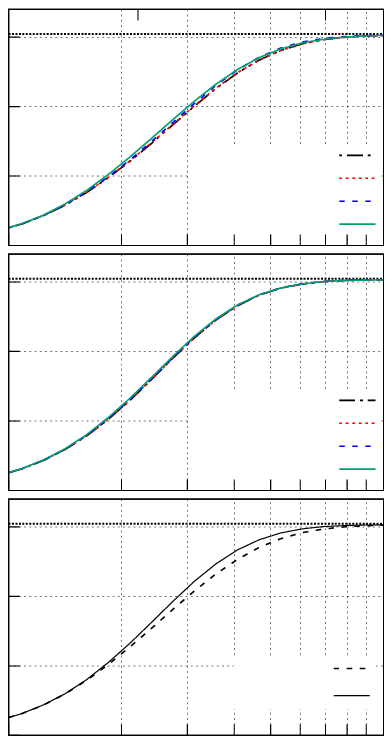}}
\end{center}
\vspace{0.2cm}
 \caption{
 \label{fig:trh_vs_neff}
 Relations between $T_{\rm RH}$ and $N_{\rm eff}$. The top- and middle panels respectively 
 show the effect of neutrino oscillation in the case without and with neutrino self-interaction, 
 while the bottom one shows the effect of neutrino self-interaction in the case with 
 $\theta_{12}$. The canonical value $N_{\rm eff} = 3.046$~\cite{Mangano_2005} 
 is also plotted with the black dotted horizontal line.
 }
\vspace{-0.2cm}
\end{figure}

Figure~\ref{fig:trh_vs_neff} shows the relation between $T_{\rm RH}$ and $N_{\rm eff}$ for the cases with 
and without neutrino self-interaction. As shown in Fig.~\ref{fig:trh_vs_neff}, the value of $N_{\rm eff}$ 
increases as $T_{\rm RH}$ becomes large, and the value is almost equal to 3.046 above 
$T_{\rm RH} \gtrsim$\ 10~MeV which is the canonical value in the standard big-bang cosmology 
with large $T_{\rm RH}$~\cite{Mangano_2005,Salas_2016}. The above threshold value of 
$T_{\rm RH}$ arises from the fact that weak reaction processes which are responsible 
for the neutrino thermalization decouple at around a temperature $T_{\rm dec}$ given by 
$\Gamma_{\rm weak}/H \sim G_F^2 T_{\rm dec}^5/(T_{\rm dec}^2/m_{\rm pl}) \sim 1$, 
{\it i.e.} $T_{\rm dec} \sim (G_F^2 m_{\rm pl})^{-\frac{1}{3}} \sim {\cal O}(1)$~MeV 
where $\Gamma_{\rm weak}$ is the thermal reaction rate of weakly-interacting particles. 
Therefore, if $T_{\rm RH}$ is larger than $T_{\rm dec}$, neutrinos have enough time 
to be fully thermalized before decoupling. In addition, it can be seen from Fig.~\ref{fig:trh_vs_neff} 
that both neutrino oscillation and neutrino self-interaction increase the value of $N_{\rm eff}$. 
This is because the production rate of $\nu_e$ is larger than that of $\nu_x$ 
($R_{\nu_e} > R_{\nu_x}$), and thereby neutrino oscillation increases the total 
production rate of neutrinos $R_{\nu,\,{\rm tot}}\, (\equiv R_{\nu_e} + R_{\nu_x})$. 
To understand this effect more quantitatively, let us assume that all neutrino species 
are almost thermalized. In this case, we can approximate the production rates of 
$\nu_e$ and $\nu_x$ as~\cite{Bell_1999}
\begin{eqnarray}
 R_{\nu_e} &\sim& C_e G_F^2 T_\gamma^5 (f_{\rm eq} - f_{\nu_e})\,,\\
 R_{\nu_x} &\sim& C_x G_F^2 T_\gamma^5 (f_{\rm eq} - f_{\nu_x})\,, 
\end{eqnarray}
where $f_{\rm eq}$ is the Fermi-Dirac distribution $f_{\rm eq} = 1/(1+\exp(p/T_\gamma))$, 
and $f_{\nu_e}$ and $f_{\nu_x}$ are distribution functions of $\nu_e$ and $\nu_x$, respectively.
Also, $C_e \sim 1.27$ is the collision coefficient for $\nu_e$ and $C_x \sim 0.92$ 
is that for $\nu_x$~\cite{Enqvist_1992}. By denoting the effect of neutrino oscillation 
at a certain time by  
$\Delta f\, \equiv f_\nu|_{\rm with\, osci} - f_\nu|_{\rm no\, osci} \equiv - \Delta f_{\nu_e} = \Delta f_{\nu_x}$, 
we can evaluate the effect of neutrino oscillation on the total production rate of neutrinos 
$\Delta R_{\nu,\,{\rm tot}} \equiv R_{\nu,\,{\rm tot}}|_{\rm with\, osci} - R_{\nu,\,{\rm tot}}|_{\rm no\, osci} 
= (R_{\nu_e}|_{\rm with\, osci} + R_{\nu_x}|_{\rm with\, osci}) - (R_{\nu_e}|_{\rm no\, osci} 
+ R_{\nu_x}|_{\rm no\, osci}) = (R_{\nu_e}|_{\rm with\, osci} - R_{\nu_e}|_{\rm no\, osci}) 
+ (R_{\nu_x}|_{\rm with\, osci} - R_{\nu_x}|_{\rm no\, osci}) 
\equiv \Delta R_{\nu_e} + \Delta R_{\nu_x}$ as follows:
\begin{eqnarray}
 \Delta R_{\nu,\,{\rm tot}} &=& \Delta R_{\nu_e} + \Delta R_{\nu_x}\,, \nonumber \\
  &\sim& G_F^2 T_\gamma^5 ( - C_e \Delta f_{\nu_e} - C_x \Delta f_{\nu_x})\,, \nonumber \\
  &=& G_F^2 T_\gamma^5 (C_e - C_x) \Delta f\,. 
\end{eqnarray}
As we can see from the expression, the quantity $\Delta R_{\nu,\,{\rm tot}}$ is larger 
than zero when $\Delta f=f_{\nu_e}-f_{\nu_x} > 0$ which holds if the reheating temperature 
is sufficiently low for neutrinos to be fully thermalized and thereby $f_{\nu_e} > f_{\nu_x}$. 
Consequently, we can see that neutrino oscillation increase the total production rate of 
neutrinos unless all neutrinos are completely thermalized. 
Neutrino self-interaction plays a role similar to neutrino oscillation. That is, they 
equilibrate abundances of neutrinos among themselves and enhance the thermalization of 
neutrinos in the same way as neutrino oscillation.

\begin{figure}[!t]
\begin{center}
\resizebox{11cm}{!}{\input{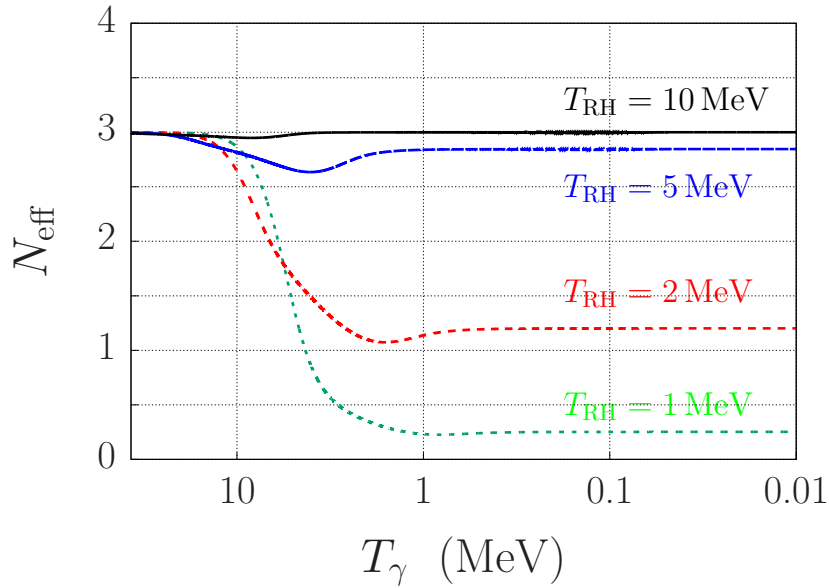}}
\end{center}
 \caption{
 \label{fig:neff_evolution}
 Time evolution of $N_{\rm eff}$ for each value of $T_{\rm RH}$. The black solid line 
 is for $T_{\rm RH} = 10$~MeV, the blue long-dashed line is for $T_{\rm RH} = 5$~MeV, 
 the red middle-dashed line is for $T_{\rm RH} = 2$~MeV, and the green short-dashed line 
 is for $T_{\rm RH} = 1$~MeV. Neutrino oscillation with $\theta_{12}$ and neutrino 
 self-interaction are considered in the calculation.
 }
\end{figure}

We can also see from Fig.~\ref{fig:trh_vs_neff} that the effect of $\theta_{13}$ is much 
smaller than that of $\theta_{12}$ or neutrino self-interaction. This is true independent 
of the neutrino mass ordering. The relative differences of effects among different mixings 
can be understood as follows. If the vacuum term ($\mathcal{H}_{\rm vac}$) dominates 
other matter terms ($\mathcal{H}_{\rm mat}$) and neutrino oscillation occur adiabatically, 
the effective transition rate of neutrinos from one flavor to another ({\it i.e.} 
$\nu_\alpha \rightarrow \nu_\beta$ where $\alpha \neq \beta$) due to neutrino oscillation 
can be written as~\cite{Foot_1996}
\begin{equation}
 \Gamma_{\rm trans} = \frac{1}{4}\sin 2\theta \,\Gamma_{\rm coll}\,,  
\end{equation}
where $\Gamma_{\rm coll}$ is the collision rate of neutrinos.  
Therefore, the value of the mixing angle solely determines how large oscillation happens 
in this case. As we can see from Eqs.~\eqref{mixing_para1}--\eqref{mixing_para3}, 
the value of $\sin^2\theta_{12}$ is almost ten times larger than that of $\sin^2\theta_{13}$. 
That is the reason that the effect of $\theta_{12}$ on the neutrino thermalization is larger 
than $\theta_{13}$ in the case of normal mass ordering. In the case of inverted mass ordering, 
neutrino oscillation proceeds via MSW resonance, and non-adiabatic effects can be important. 
As for this point, authors in Ref.~\cite{Johns_2016} evaluated the adiabaticity of the MSW 
resonance and concluded that the non-adiabatic effects are negligible.
Therefore, an efficient oscillation should occur when a large population of neutrinos 
go though the resonance even if we adopt the reactor neutrino mixing $\theta_{13}$. 
Since the MSW resonance happens at around a temperature of $T_c \sim 5$~MeV 
for neutrino with $p = \langle p \rangle \sim 3.15\,T_\gamma$, we can expect 
larger oscillation effects in the case of $\theta_{13}$~(IO) at $T_{\rm RH} < T_c$, 
which in fact can be seen in Fig.~\ref{fig:trh_vs_neff}.

\begin{figure}[!t]
\vspace{-0.3cm}
\begin{center}
\resizebox{20cm}{!}{\ \ \ \ \ \ \ \input{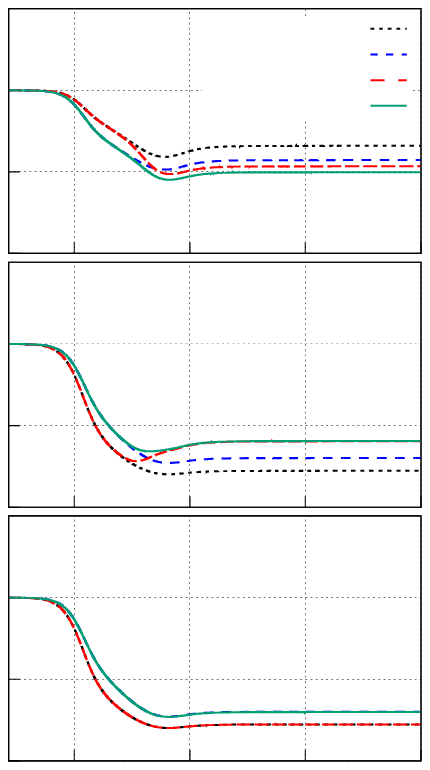}}
\end{center}
\vspace{0.5cm}
 \caption{
 \label{fig:neff_evolution2}
 Time evolution of $N_{{\rm eff},\,\nu_\alpha}$ 
 for $T_{\rm RH} = 2$~MeV. $N_{\rm eff,\, \nu_e}$ (top panel) is for $\nu_e$, 
 $N_{{\rm eff},\, \nu_x}$ (middle panel) is for $\nu_x$ and $N_{{\rm eff},\,\nu_{\rm sp}}$ 
 (bottom panel) is for $\nu_{\rm sp}$. In the red long-dashed and green solid lines, 
 neutrino oscillation with $\theta_{12}$ is taken into account, 
 and neutrino self-interaction is considered in the blue middle-dashed and green 
 solid lines. In the figure of $N_{{\rm eff},\,\nu_{\rm sp}}$, the black short-dashed 
 (blue middle-dashed) and red long-dashed (green solid) lines are overlapping.
 }
\end{figure}

Figure~\ref{fig:neff_evolution2} shows the time evolution of $N_{\rm eff}$, and 
Fig.~\ref{fig:neff_evolution2} is the same as Fig.~\ref{fig:neff_evolution}, 
but for the contribution for each neutrino species $N_{{\rm eff},\,{\nu_\alpha}}$ 
($\alpha = e,\,x,\, {\rm sp}$). 
Since the final abundance of neutrinos does not depend on the condition before reheating, 
we assume that neutrinos have thermal spectra ({\it i.e.} Fermi-Dirac distributions) at the 
initial time. We can see from Fig.~\ref{fig:neff_evolution} that the value of $N_{\rm eff}$ 
decreases until $T_{\rm RH}$ is realized. This is due to the entropy production from decays 
of the massive particles. The value of $N_{\rm eff}$ then increases at $T_\gamma < T_{\rm RH}$ 
until neutrinos are decoupled from other particles at around a few MeV. This corresponds 
to the upturn behavior in the evolution of $N_{\rm eff}$. In addition, we can see from the 
evolution of $N_{{\rm eff},\,{\nu_\alpha}}$ in Fig.~\ref{fig:neff_evolution2} that 
neutrino oscillation becomes effective at around a temperature of a few MeV, whereas 
neutrino self-interaction becomes effective at higher temperature. The former is because, 
as we have discussed in the previous section, neutrino oscillation with the solar neutrino 
mixing ($\delta m_{12}^2, \theta_{12}$) are effective when the photon temperature is lower 
than $T_c \sim 3$~MeV. The latter is due to the reason that the reaction rates of neutrino 
self-interactions monotonically increase with the photon temperature.

\begin{figure}[!t]
\vspace{-1.8cm}
\resizebox{16cm}{!}{\ \ \input{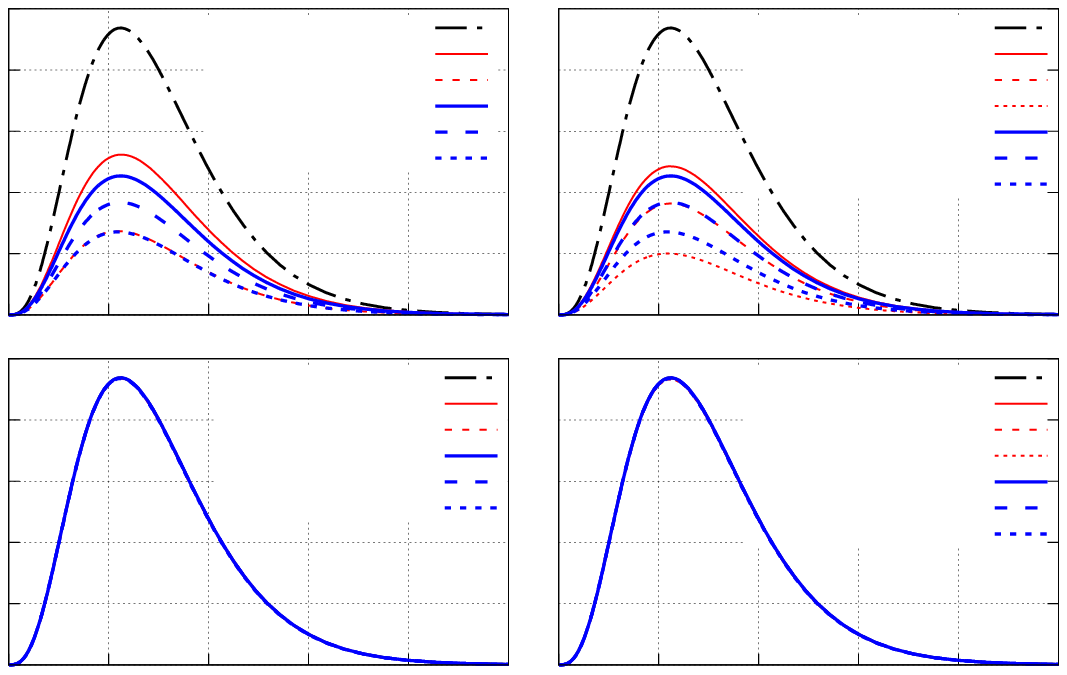}}
\vspace{0.2cm}
 \caption{
 \label{fig:nu_spectrum}
 Final energy spectra of neutrinos. 
 The horizontal axis is the neutrino energy $p$ divided by the photon temperature 
 $T_\gamma$. The vertical axis is the differential energy spectrum of neutrinos. 
 In the figure, neutrino self-interaction is considered in the left column, 
 whereas neutrino oscillation with $\theta_{12}$ is considered 
 in the right column. We consider $\theta_{12}$ in the case 
 with neutrino oscillation in the left column. The thermal spectrum is plotted 
 with the black dot-dashed line. In the top-left panel, the thin red short-dashed 
 and blue thick short-dashed lines are overlapping, while the red thin long-dashed 
 and blue thick long-bashed lines are overlapping in the top-right panel. 
 Also, all plots are almost overlapping in the case of $T_{\rm RH} = 10$~MeV.
 }
\end{figure}

The role of neutrino oscillation and neutrino self-interaction is shown in 
Fig.~\ref{fig:nu_spectrum}, where we plot final energy spectra of neutrinos for 
the cases with and without neutrino oscillation or neutrino self-interaction. 
The energy spectra are evaluated at $T_\gamma \sim 10^{-2}$~MeV which corresponds 
to the epoch well after electron annihilation. In the case of $T_{\rm RH} = 2$~MeV, 
both of these effects decrease the difference in neutrino abundances. On the other hand, 
if $T_{\rm RH}$ is large enough ({\it e.g.}~$T_{\rm RH} = 10$~MeV), neutrinos are 
almost completely thermalized well before decoupling. Therefore, neutrino oscillation 
or neutrino self-interaction plays no role in the final abundance of neutrinos. 

\begin{figure}[!t]
\vspace{-0.5cm}
\begin{center}
\resizebox{24.0cm}{!}{\ \ \ \ \ \ \ \ \ \ \ \ \ \ \ \ \ \ \ \ \ \ \ \ \ \input{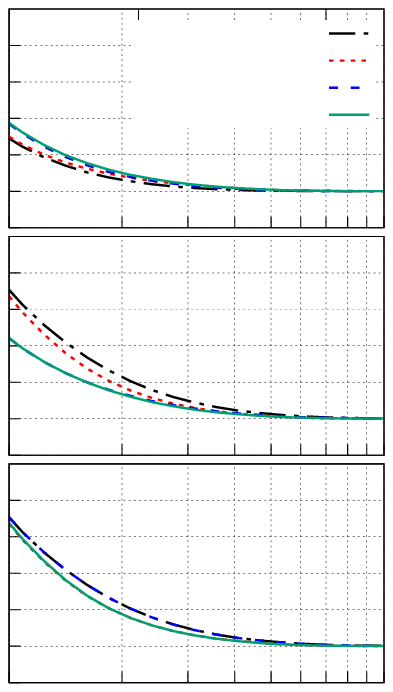}}
\end{center}
\vspace{0.5cm}
 \caption{
 \label{fig:trh_vs_Rdist}
 Dependence of the mean energy of $\nu_e$ (top panel), $\nu_x$ (middle panel) 
 and $\nu_{{\rm sp}}$ (bottom panel) on $T_{\rm RH}$. The vertical axis is the distortion 
 parameter $R_{\rm dist}$ for each neutrino species. $R_{\rm dist}$ = 1 corresponds to 
 the thermal spectrum. In the middle panel, the blue long-dashed and green solid lines 
 are overlapping. Also, in the bottom panel, the black dot-dashed (red short-dashed) 
 and blue long-dashed (green solid) lines are overlapping. We consider 
 $\theta_{12}$ in the case with neutrino oscillation.
 }
\end{figure}

Figure~\ref{fig:trh_vs_Rdist} shows the dependence of the mean energy of 
$\nu_e$, $\nu_x$ and $\nu_{\rm sp}$({\it i.e.}~$\rho_{\nu_\alpha}/n_{\nu_\alpha}$) 
on $T_{\rm RH}$. The quantity $R_{\rm dist}$ on the vertical axis was introduced 
to measure the distortion in the final energy spectrum of neutrinos in 
Ref.~\cite{Kawasaki_2000}, and it is defined as 
\begin{equation}
 R_{\rm dist} = \frac{1}{3.15\,T_{\nu,\,{\rm eff}}}\frac{\rho_\nu}{n_\nu} \,, 
\label{eq:R_dist}
\end{equation}
where $T_{\nu,\,{\rm eff}}$ is the effective temperature of neutrinos 
$T_{\nu,\,{\rm eff}} = [\frac{4\pi^2}{3\zeta(3)}n_\nu]^{1/3}$. As we can see 
from this definition, $R_{\rm dist} = 1$ corresponds to the thermal spectrum, 
and $R_{\rm dist} > 1$ indicates a larger mean energy. It can be seen from 
Fig.~\ref{fig:trh_vs_Rdist} that the value of $R_{\rm dist}$ increases as $T_{\rm RH}$ 
becomes smaller. This is because neutrinos are only produced from the annihilation 
of electrons $e^- + e^+ \rightarrow \nu_\alpha + \bar{\nu}_\alpha$, and 
neutrinos in the final state therefore have an energy larger than twice 
the electron mass. Thus, if neutrinos are mainly produced when the 
electron mass is not negligible, and the equilibration process 
$e^\pm + \nu_\alpha \rightarrow e^\pm + \nu_\alpha$ is not effective as 
in the case of $T_{\rm RH} \lesssim {\cal O}(1)$~MeV, $R_{\rm dist}$ 
becomes larger than unity.
In addition, since $\nu_e$ scatter with electrons stronger than $\nu_x$ 
and $\nu_{\rm sp}$ due to the charged-current interaction, the energy distribution 
of $\nu_e$ is closer to the thermal spectrum. That is the reason that the relation
$R_{\rm dist,\, \nu_e} < R_{{\rm dist},\, \nu_x}$, 
$R_{\rm dist,\, \nu_e} < R_{{\rm dist},\, \nu_{\rm sp}}$ 
holds for a sufficiently small $T_{\rm RH}$. Furthermore, we can see from 
Fig.~\ref{fig:trh_vs_Rdist} that both neutrino oscillation and neutrino 
self-interaction increase $R_{{\rm dist},\, \nu_e}$, while decrease $R_{{\rm dist},\, \nu_x}$. 
This is because neutrino oscillation and neutrino self-interaction equilibrate 
the neutrino abundances of different flavors as shown in Fig.~\ref{fig:nu_spectrum}. 
The reason is as follows: If the final distribution function of neutrinos is changed 
by a factor of $\kappa$ ({\it i.e.}~$f_\nu \mapsto \kappa f_\nu$ where $\kappa < 1$ 
for $\nu_e$ and $\kappa > 1$ for $\nu_x$) due to neutrino oscillation or 
neutrino self-interaction, then the distortion parameter should be also modified by 
\begin{equation}
 \widetilde{R}_{\rm dist}/R_{\rm dist}  
  = \frac{\kappa \rho_\nu / \kappa n_\nu}{3.15 (\kappa^{1/3} T_{\nu,\,{\rm eff}})}
  \left(\frac{\rho_\nu / n_\nu}{3.15 T_{\nu,\,{\rm eff}}}\right)^{-1} = \kappa^{-1/3}\,,
\end{equation}
where $\widetilde{R}_{\rm dist}$ and $R_{\rm dist}$ are distortion parameters 
for the cases with and without effects of neutrino oscillation or neutrino self-interaction, 
respectively. Therefore, these effects increase $R_{\rm dist,\, \nu_e}$ and decrease 
$R_{{\rm dist},\, \nu_x}$ as long as neutrino oscillation or neutrino self-interaction is  
effective. In addition, since the reaction rate of neutrino self-interaction strongly 
depends on the number density of neutrinos, its effect on $R_{\rm dist}$ becomes small 
more rapidly than neutrino oscillation as $T_{\rm RH}$ decreases. These effects on 
$R_{\rm dist}$ can be estimated by comparing neutrino distribution functions in 
Fig.~\ref{fig:nu_spectrum} and is consistent with the results in Fig.~\ref{fig:trh_vs_Rdist}. 
On the other hand, since $\nu_{\rm sp}$ does not mix with other flavor of neutrinos, 
they are only affected by neutrino self-interaction.

In the next section, we discuss the light element abundances created in the process 
of BBN taking our computed neutrino thermalization into account.

\section{\label{Sec:results_BBN}Big Bang Nucleosynthesis}
As mentioned in Sec.~\ref{Sec:dynamics}, incomplete thermalization of
neutrinos affects the dynamics of the standard BBN. In this section,
we explain the role of neutrinos in the production process of light
elements and show our results of BBN obtained by assuming
$T_{\rm RH} \sim {\cal O}(1)$~MeV.

\subsection{Formulation of BBN}
We have seen in the previous section that the late-time entropy production 
due to decays of $X$ induces the incomplete thermalization of neutrinos 
before decoupling. Since neutrinos take part in the weak reaction processes,
\begin{align}
 n &\leftrightarrow p + e^- + \bar{\nu}_e\,,\label{Eq:np_rate1}\\
 e^+ + n &\leftrightarrow p + \bar{\nu}_e\,,\label{Eq:np_rate2}\\
 \nu_e + n &\leftrightarrow p + e^-\,,\label{Eq:np_rate3}
\end{align}
which interchange ambient neutrons and protons with each other,
non-thermal spectra of neutrinos significantly change the freeze-out
value of the neutron-to-proton ratio
$(n/p)_{\rm f} \equiv (n_n/n_p)_{T = T_f}$ where $n_n$ and $n_p$ are
the number density of neutrons and protons, respectively, whereas
$T_f$ is the freeze-out temperature of the 
processes~\eqref{Eq:np_rate1}--\eqref{Eq:np_rate3}. As described later 
in this section, the theoretical values of light element abundances 
are very sensitive to the neutron-to-proton ratio before BBN. 
Therefore, theoretical predictions of the standard BBN should be 
modified in the Universe with small $T_{\rm RH}$. Since the predictions 
of standard BBN is well consistent with the observational values, 
we can constrain $T_{\rm RH}$ by requiring that the late-time entropy 
production does not spoil the current success of the standard BBN.

In the case where the massive particles have a hadronic branching ratio, 
there are additional neutron-proton interchanging processes other 
than~\eqref{Eq:np_rate1}--\eqref{Eq:np_rate3} via strong interactions 
caused by injected hadrons $N + H \leftrightarrow N' + H'$ 
where $N$ and $N'$ are nucleons, and $H$ and $H'$ are mesons or baryons. 
If the hadronic branching ratio is large enough, the hadronic processes 
dominantly affect the neutron-to-proton ratio, which result in different 
light element abundances compared to the case of the 100\% radiative decays 
of the massive particles~\cite{Reno_1988,Kawasaki_2000}. In the current study, 
we consider the hadronic processes involving pions ($\pi^\pm$) and nucleons 
($n,\,\bar{n},\,p,\,\bar{p}$) which are injected from hadronic decays 
of the massive particles. The energetic hadrons produced in the decay of 
the massive particles are instantaneously stopped by Coulomb scattering 
with background electrons/positrons or inverse-Compton like scattering 
with background photons~\cite{Reno_1988,Kohri_2001,Kawasaki_2005}. 
Therefore, the hadrons affecting neutron-proton inter-conversions are 
thermalized, and we can use thermal cross sections for the calculation. 
As for the hadronic cross sections, we adopt those given in Table.1 of 
Ref.~\cite{Reno_1988} for the mean values and assume $30\%$ experimental 
error in each cross section for a conservative treatment 
(see also Refs.~\cite{Kawasaki_2000,Pospelov_2010}).

In order to follow the evolution of light element abundances, we solve 
the Boltzmann equations of light elements using the Kawano code~\cite{Kawano_1992}. 
Since some of the nuclear reaction rates in the code are already outdated, 
we replace them with the latest ones (see Ref.~\cite{Kawasaki_2018} 
for more information). In addition, we rewrite some equations in the code 
to allow for the late-time entropy production accompanied by the decays of $X$. 
Moreover, since the free neutron decay ({\it i.e.} the forward process of~\eqref{Eq:np_rate1}) 
continues even after the other weak processes of~\eqref{Eq:np_rate1}--\eqref{Eq:np_rate3} 
decoupled at $T_\gamma \sim T_f$, the value of the neutron-to-proton ratio just 
before BBN depends on the lifetime of neutrons (see {\it e.g.}~\cite{Steigman_2007}). 
In the current study, we use the value of the neutron lifetime $\tau_n$ 
reported in Ref.~\cite{PDG2018}:
\begin{equation}
 \tau_n = 880.2 \pm 1.0~{\rm sec}\ \ (68\%~{\rm C.L.})\ \,.
\end{equation}

As for the observational values of light elements, we adopt the primordial 
mass fraction of helium $^4$He, $Y_p$, reported in Ref.~\cite{Aver_2015}:
\begin{equation}
 Y_p = 0.2449 \pm 0.0040\ \ (68\%~{\rm C.L.}) \,, \label{Eq:obshelium}
\end{equation}
whereas for the observational value of primordial abundance of deuterium D, 
we adopt the latest value reported in Ref.~\cite{Zavarygin_2018}:
\begin{equation}
 {\rm D/H} = (2.545 \pm 0.025) \times 10^{-5}\ \ (68\%~{\rm C.L.})\,.\label{Eq:obsdeuterium}
\end{equation}
%

\begin{figure}[!t]
\vspace{1cm}
\begin{center}
\resizebox{20cm}{!}{\ \ \ \ \ \ \ \ \ \ \ \ \ \ \input{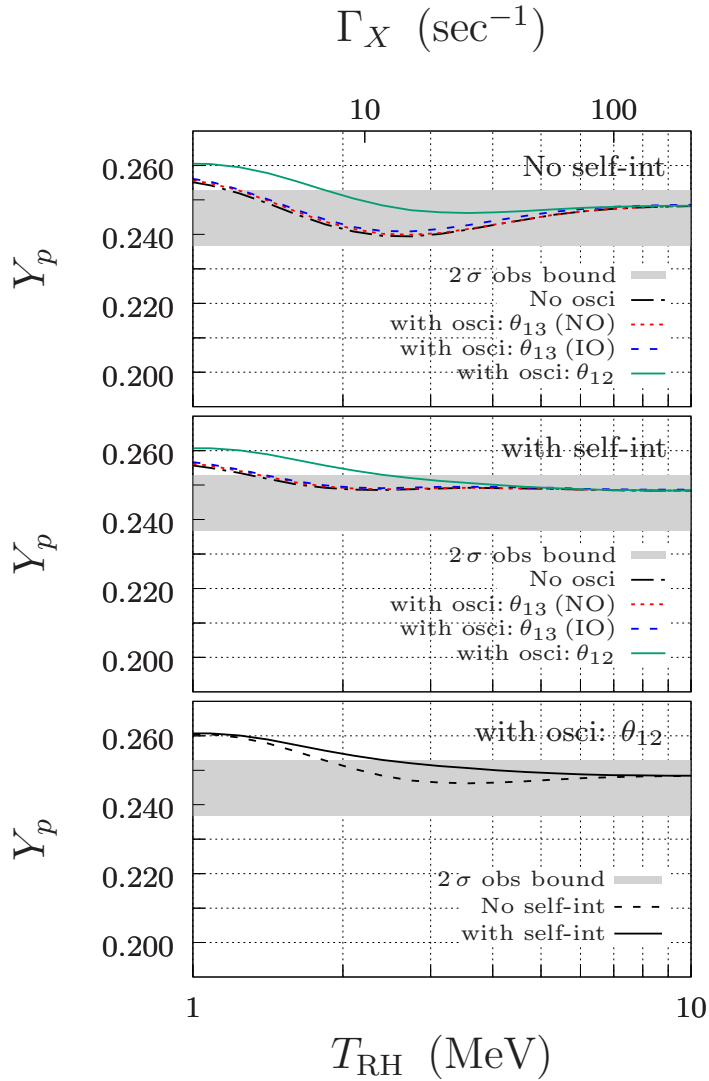}}
\end{center}
\vspace{0.5cm}
 \caption{
 \label{fig:trh_vs_Yp}
 Relations between $T_{\rm RH}$ and $Y_p$ in the case of the 100\% radiative 
 decays of $X$. We adopt $\eta_B = 6.13 \times 10^{-10}$ in the figure. 
 The top- and middle panels show the effect of neutrino oscillation for the cases 
 without and with neutrino self-interaction, respectively. The black dot-dashed line 
 is for the case without neutrino oscillation, the red short-dashed line is for the case 
 with $\theta_{13}$ (NO), the blue long-dashed line is for the case with $\theta_{13}$ (IO), 
 and the green solid line is for the case with $\theta_{12}$. The bottom panel shows the 
 effect of neutrino self-interaction when we consider neutrino oscillation with 
 $\theta_{12}$. The black dashed- and solid lines are for the cases 
 without and with neutrino self-interaction, respectively. The gray-shaded region 
 corresponds to the $2\sigma$ observational bound.
 }
\end{figure}

\begin{figure}[!t]
\vspace{1cm}
\begin{center}
\resizebox{20cm}{!}{\ \ \ \ \ \ \ \ \ \ \ \ \ \ \input{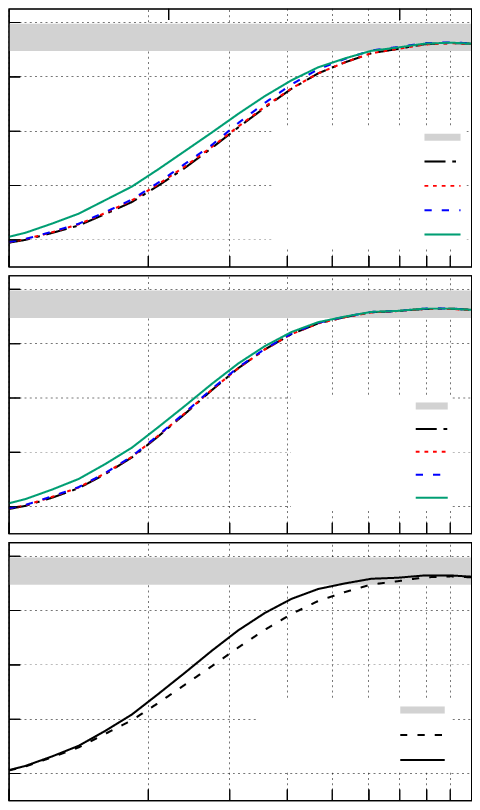}}
\end{center}
\vspace{0.2cm}
 \caption{
 \label{fig:trh_vs_D}
 Same as Fig.~\ref{fig:trh_vs_Yp}, but for D/H.
 }
\end{figure}

\subsection{Results of BBN: Radiative decay}
First we show the results of radiative decay, {\it i.e.} the hadronic
branching ratio Br $= 0$. In this case, photons and charged leptons
emitted from the decay of $X$ are instantaneously thermalized via
electromagnetic force, and results of neutrino thermalization and 
BBN are independent of the mass of $X$, $m_X$. 
In Figs.~\ref{fig:trh_vs_Yp} and \ref{fig:trh_vs_D}, the relation
between $T_{\rm RH}$ and D/H and $Y_p$ are shown, respectively.  
We assume the 100\% radiative decays of the massive particles in these
figures. The baryon-to-photon ratio $\eta_B$ is the only free
parameter in the standard BBN. In the low-reheating-temperature
Universe, a baryon number is diluted by the entropy production due to
the decays of the massive particles, and hence $\eta_B$ is
decreased by many orders of magnitude.  Therefore, we set the large
initial value of $\eta_B$ so that the final value of
$\eta_B$ is consistent with observations of light elements. To
plot Figs.~\ref{fig:trh_vs_Yp} and \ref{fig:trh_vs_D}, we fix
the final value of $\eta_B$ to the median value reported by
Planck collaboration~\cite{Planck_2018}:
\begin{equation}
 \eta_B = 6.13 \times 10^{-10}\,. 
\label{baryon_asym} 
\end{equation}
Since almost all neutrons are processed into $^4$He which is the most 
stable among light elements, the primordial mass fraction of $^4$He 
can be written as $Y_p \equiv \rho_{\rm ^4He}/\rho_B 
\sim 2/\{1+(n/p)_{\rm BBN}^{-1}\} \sim 0.25$, where $(n/p)_{\rm BBN} 
\equiv (n/p)_{\rm f}\,e^{-t/\tau_n}$ is the neutron-to-proton ratio 
just before deuterium bottleneck opens ({\it i.e.} $T_\gamma \sim 0.08$~MeV 
and $t \sim 200$~sec), and the last approximation holds in the standard 
big-bang cosmology where $(n/p)_{\rm BBN} \sim 1/7$~\cite{Steigman_2007}.
Therefore, the value of $(n/p)_{\rm BBN}$ almost entirely determines the 
final abundance of $^4$He. As for the final abundance of D, the value of 
$N_{\rm eff}$ is also important because it is related to the Hubble parameter 
(see Eqs.~\eqref{Eq:rhotot}, \eqref{Eq:Friedmann} and \eqref{Eq:neff_def}) 
and determines when each light element departs from the nuclear statistical 
equilibrium~\cite{Esmailzadeh_1990,Smith_1993}.~\footnote{
  We can intuitively understand the dependence of the D abundance 
  on the expansion rate of the Universe in the BBN epoch by focusing on 
  the binding energy of D and $^4$He, {\it i.e.} $B_{\rm D} \sim 2.22$~MeV 
  and $B_{^4{\rm He}} \sim 28.3$~MeV, and the freeze-out temperature of 
  the destroying reactions of D. 
  Since the binding energy of $^4$He is much larger than that of D, D should 
  burn into $^4$He (via mass-3 elements, T and $^3$He) as long as the destroying 
  reactions of D such as DD and DT fusions are effective. For this reason, 
  a small value of $N_{\rm eff}$ (or equivalently a small expansion rate $H$), 
  attained in the low reheating temperature cases, delays a decoupling of the 
  destroying reactions, and hence a smaller abundance of D remains unburnt. 
  That is the reason that a large expansion rate in the BBN epoch leads to 
  a large abundance of D and vice versa.
  } 
In addition, as we can see from Figs.~\ref{fig:trh_vs_Yp} and \ref{fig:trh_vs_D}, 
the influences of neutrino oscillation and self-interaction on light element 
abundances are similar, and both of these effects increase $Y_p$ and D/H. 
In order to understand the numerical results on light element abundances, 
next we focus on the dynamics of the freeze-out of the neutron-to-proton ratio.

Since nucleons are always non-relativistic, $(n/p)_{\rm f}$ can be expressed 
with the freeze-out temperature if $T_{\rm RH}$ is MeV scale as 
$(n/p)_{\rm f} \sim \exp(-Q/T_{\rm f})$ where $Q \equiv m_n - m_p \sim 1.3$~MeV 
is the mass difference of nucleons. We note that $T_{\rm f}$ is determined by 
the relative values of the neutron-proton inter-converting weak reaction rates 
$\Gamma_{np}$ and the Hubble parameter $H$ and is roughly given by 
$\Gamma_{np}(T_{\rm f})/H(T_{\rm f}) \sim 1$. In the low-reheating-temperature 
Universe, the total energy density is smaller than that in the standard big-bang 
cosmology under the same photon temperature due to the incomplete thermalization 
of neutrinos.~\footnote{
  We recall the reader that $T_\gamma$ determines when light element abundances 
  are to be created since the reaction process $p + n \rightarrow D + \gamma$ 
  responsible for the deuterium production is the first step of BBN, 
  and its backward reaction rate depends on $T_\gamma$. For this reason, 
  the values of $\Gamma_{np}$ and $H$ should not be characterized by the 
  cosmic time but rather by $T_\gamma$. That is the reason that a larger value 
  of $N_{\rm eff} \propto \rho_\nu/\rho_\gamma$ leads to a larger expansion rate 
  of the Universe at the epoch of BBN.
  } 
Therefore, the expansion rate of the Universe is also small in the scenario, 
and this effect delays the decoupling of the processes~\eqref{Eq:np_rate1}--\eqref{Eq:np_rate3} 
and thereby decreases $(n/p)_{\rm f}$. 

The influence is not only in the Hubble parameter $H$ but also in the 
reaction rates $\Gamma_{np}$. Specifically, the reaction rate of the 
processes~\eqref{Eq:np_rate1}--\eqref{Eq:np_rate3} can be written as~\cite{Kawasaki_2000}
\begin{eqnarray*}
     \Gamma_{n \to p e^- \bar{\nu}_e}& = & K\int_0^{Q-m_e}dp'\left[\sqrt{(p'-Q)^2-m_e^2}(Q-p')
     \frac{{p'}^2}{1+e^{(p'-Q)/T_{\gamma}}} \left(1-f_{\nu_e}(p')\right) \right] \,,\\  
     \Gamma_{ne^+ \to p \bar{\nu}_e} & = & K \int_{Q+m_e}^{\infty}dp'\left[\sqrt{(p'-Q)^2-m_e^2}(p'-Q)
     \frac{{p'}^2}{e^{(p_{\nu_e - Q})/T_{\gamma}}+1} \left( 1-f_{\nu_e}(p') \right) \right] \,,\\ 
     \Gamma_{n \nu_e \to p e^-} & = & K \int_0^{\infty}dp'\left[\sqrt{(p'+Q)^2-m_e^2}(p'+Q)
     \frac{{p'}^2}{1+e^{-(p'+Q)/T_{\gamma}}} f_{\nu_e}(p')\right] \,,\\ 
     \Gamma_{pe^-\bar{\nu}_e \to n} & = & K \int_{0}^{Q-m_e}dp'\left[\sqrt{(p'-Q)^2-m_e^2}(Q-p')
     \frac{{p'}^2}{e^{-(p_{\nu_e - Q})/T_{\gamma}}+1} f_{\nu_e}(p') \right] \,,\\ 
     \Gamma_{pe^- \to n \nu_e} & = & K \int_{0}^{\infty}dp'\left[\sqrt{(p'+Q)^2-m_e^2}(Q+p')
     \frac{{p'}^2}{e^{(p_{\nu_e + Q})/T_{\gamma}}+1} \left( 1-f_{\nu_e}(p') \right)\right]\,,\\
     \Gamma_{p\bar{\nu}_e \to ne^+} & = & K \int_{Q+m_e}^{\infty}dp'\left[\sqrt{(p'-Q)^2-m_e^2}(Q-p')
     \frac{{p'}^2}{1+e^{-(p_{\nu_e - Q})/T_{\gamma}}} f_{\nu_e}(p') \right]\,,
\end{eqnarray*}
where $m_e$ is the electron mass, and $K \sim (1.636 \tau_n)^{-1}$ 
is a normalization factor whose value is determined by the neutron lifetime 
$\tau_n$. Of these reaction rates, some depend on $f_{\nu_e}$ and 
others on $1-f_{\nu_e}$. 
In the low-reheating-temperature Universe, the neutrino abundance of each 
flavor is smaller than the case of the standard big-bang cosmology. 
Therefore, by denoting the reduction of $f_{\nu_e}$ due to the incomplete 
thermalization of neutrinos by $\Delta f_{\nu_e}$, the following relation 
holds for sufficiently small $T_{\rm RH}$ ($< T_{\rm dec} \sim {\cal O}(1)$~MeV):   
\begin{equation}
 \left|\frac{\Delta f_{\nu_e}}{(1-f_{\nu_e})}\right| \ll 
  \left|\frac{\Delta f_{\nu_e}}{f_{\nu_e}}\right|
  \qquad {\rm for } \ \  f_{\nu_e} \ll 0.5 \,.    
  \label{Eq:np_rate_cond}
\end{equation}
As a result, with such a small value of $T_{\rm RH}$, the total reaction rate 
$\Gamma_{np} \equiv \Gamma_{n \to p e^- \bar{\nu}_e} + \Gamma_{ne^+ \to p \bar{\nu}_e} 
+ \Gamma_{n \nu_e \to p e^-} + \Gamma_{pe^-\bar{\nu}_e} + \Gamma_{pe^- \to n \nu_e} 
+ \Gamma_{p\bar{\nu}_e \to ne^+}$ becomes smaller than that of the standard 
big-bang cosmology as written in Ref.~\cite{Kawasaki_2000}. Therefore, this 
effect accelerates the decoupling of the processes~\eqref{Eq:np_rate1}--\eqref{Eq:np_rate3} 
and thereby increases $(n/p)_{\rm f}$. Consequently, the relative magnitude 
of these two opposite contributions determine the net effect of incomplete 
thermalization of neutrinos on $(n/p)_{\rm f}$.
 
As described in the previous section, neutrino oscillation and self-interaction 
slightly enhance the neutrino thermalization and increase the total energy density 
of neutrinos and hence $N_{\rm eff}$. As a result, the Hubble expansion rate 
increases due to these effects. In addition, since the $\nu_e$ abundance 
is decreased by the conversion $\nu_e \rightarrow \nu_x$, and only $\nu_e$
take part in the reaction processes~\eqref{Eq:np_rate1}--\eqref{Eq:np_rate3}, 
$\Gamma_{np}$ decreases by considering these effect. 
Therefore, neutrino oscillation and self-interaction always play a role 
in delaying the freeze-out of neutron-to-proton ratio and increasing 
$T_{\rm f}$ and $(n/p)_{\rm f}$, which leads to larger values of 
$Y_p$ and D/H.

As is the case for $N_{\rm eff}$ (see Fig.~\ref{fig:trh_vs_neff}), we can also 
see from Figs.~\ref{fig:trh_vs_Yp} and \ref{fig:trh_vs_D} that the impact 
of the solar neutrino mixing ($\delta m^2_{12},\theta_{12}$) is much larger than 
that of the reactor neutrino mixing ($\delta m^2_{13},\theta_{13}$) independent 
of the neutrino mass ordering. Therefore, the effective two-flavor mixing with 
($\delta m^2_{12}, \theta_{12}$) gives a good approximation to the full 
three-flavor neutrino mixings. For this reason, we hereafter only consider 
($\delta m^2_{12}, \theta_{12}$) in the case with neutrino oscillation.

To obtain the observational constraint on $T_{\rm RH}$, we perform a
Monte-Carlo calculation of BBN and $\chi^2$ analysis at each point on
the grids of $\eta_B$ and $T_{\rm RH}$ assuming observational
values for $Y_p$ (Ref.~\cite{Aver_2015}) and D/H
(Ref.~\cite{Zavarygin_2018}).~\footnote{
  As written in {\it e.g.}~\cite{Hannestad_2004,Hamann_2011}, 
  it is technically incorrect to adopt the CMB bound 
  $\eta_B = (6.13 \pm 0.04) \times 10^{-10}$ reported by 
  the Planck collaboration~\cite{Planck_2018} as a prior of BBN 
  because the recombination process depends on the values of
  $N_{\rm eff}$ and $Y_p$, and there are correlations between
  $\eta_B$ and these quantities. In other words, CMB is 
  not independent from the neutrino thermalization and BBN. 
  In Ref.~\cite{Planck_2018}, they adopt the canonical value
  $N_{\rm eff} = 3.046$~\cite{Mangano_2005} and $Y_p$ calculated 
  by assuming the standard BBN, which are not necessarily realized 
  in the low-reheating-temperature Universe.
  }  
In the Monte-Carlo calculation, we assume that the reaction rates in 
the standard BBN, the hadronic reaction rates and the neutron lifetime 
follow Gaussian distribution and propagate their reported errors to 
obtain theoretical uncertainties on the light element abundances.  
Since an allowed region is defined by a parameter space where theoretical 
abundances of light elements explain each observational value, 
we give the lower bound on $T_{\rm RH}$ combining $\chi^2$ values of 
both D/H and $Y_p$:~\footnote{
  There remains a long-standing problem in the standard BBN that 
  the theoretical prediction of the $^7$Li abundance is approximately 
  three times larger than that of the observational value if we input 
  the value of the baryon-to-photon ratio from CMB into the calculation 
  of BBN (see {\it e.g.}~\cite{PDG2018}). Therefore, we refrain from 
  using the $^7$Li abundance to constrain $T_{\rm RH}$ in the current study.
  }
\begin{equation}
 \chi^2_{{\rm D/H}\, + \,Y_p} \equiv \chi^2_{\rm D/H} + \chi^2_{Y_p} 
  = \frac{\{{\rm (D/H)}_{\rm th}(\eta_B, T_{\rm RH})-{\rm (D/H)}_{\rm obs}\}^2}
  {\sigma^2_{\rm D,\,th}(\eta_B, T_{\rm RH}) + \sigma^2_{\rm D,\,obs}} 
  + \frac{\{Y_{p,\, {\rm th}}(\eta_B, T_{\rm RH})-Y_{p,\, {\rm obs}}\}^2}
  {\sigma^2_{Y_p,\, {\rm th}}(\eta_B, T_{\rm RH}) + \sigma^2_{Y_p,\, {\rm obs}}} \,,
\end{equation}
where $\chi^2_{\rm D/H}$ and $\chi^2_{Y_p}$ are $\chi^2$ values of D/H and $Y_p$, 
respectively. Also, $\sigma_{i,{\rm\,th}}$ and $\sigma_{i,{\rm \,obs}}$ where 
$i$ = D/H and $Y_p$ are respectively the theoretical and observational 
1$\sigma$ variance of each light element abundance. 

\begin{figure}[!t]
\begin{center}
\resizebox{9.5cm}{!}{\hspace{-0.9cm}\input{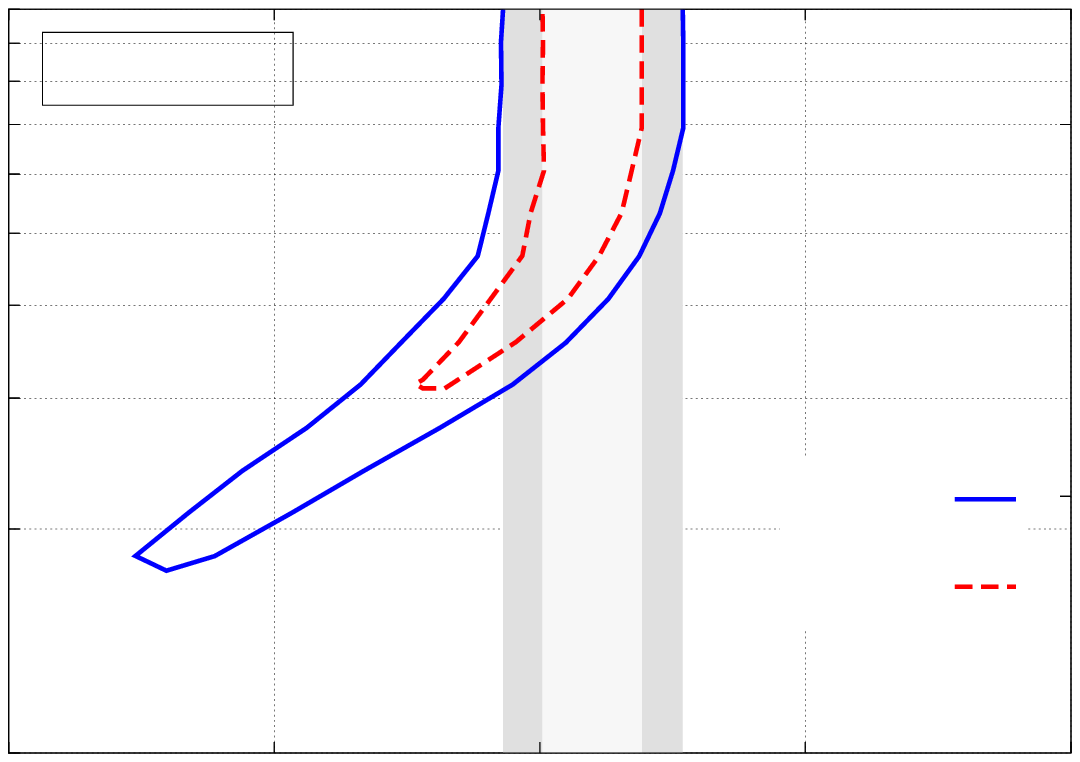}}\vspace{-1cm}
\resizebox{14.0cm}{!}{\ \ \ \ \ \ \ \input{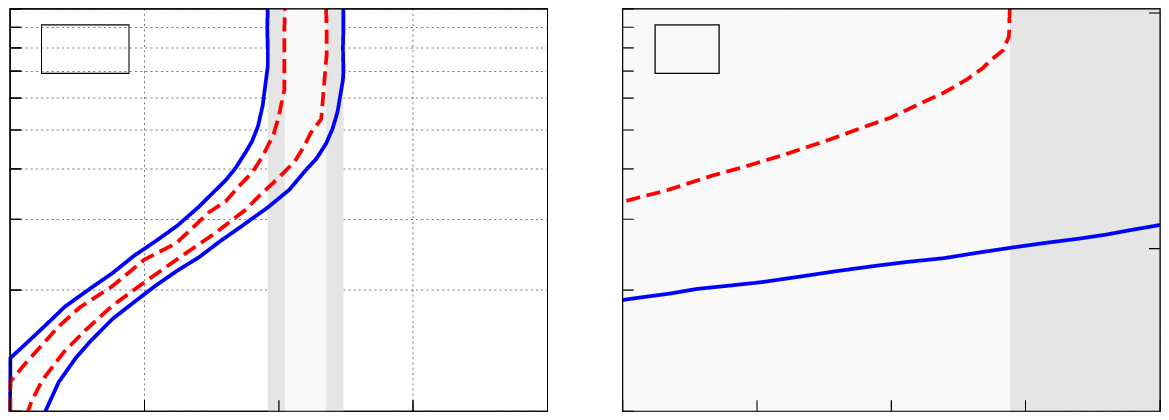}}
\end{center}
\vspace{-0.8cm}
 \caption{
 \label{fig:allowed_region_radiative}
 Allowed region in the ($\eta_B$, $T_{\rm RH}$) plane in the case of the 100\% 
 radiative decays of $X$. The 95\%~(68\%)~C.L. contour is plotted with the blue 
 solid (red dashed) line. The outside of the small region surrounded by the blue 
 solid (red dashed) line is excluded at 95\%~(68\%)~C.L. The constraint
 on $\eta_B$ at 95\% C.L. (68\% C.L.) in the case of 
 the standard BBN is also shown as the dark (light) shaded region. 
 The top panel shows the allowed region in terms of both $Y_p$ and D/H, whereas 
 the bottom-left and bottom-right panels show those of D/H and of $Y_p$, respectively. 
 Neutrino oscillation and neutrino self-interaction are considered in the calculation.
 }
\end{figure}

\begin{figure}[!t]
\begin{center}
\resizebox{11cm}{!}{\input{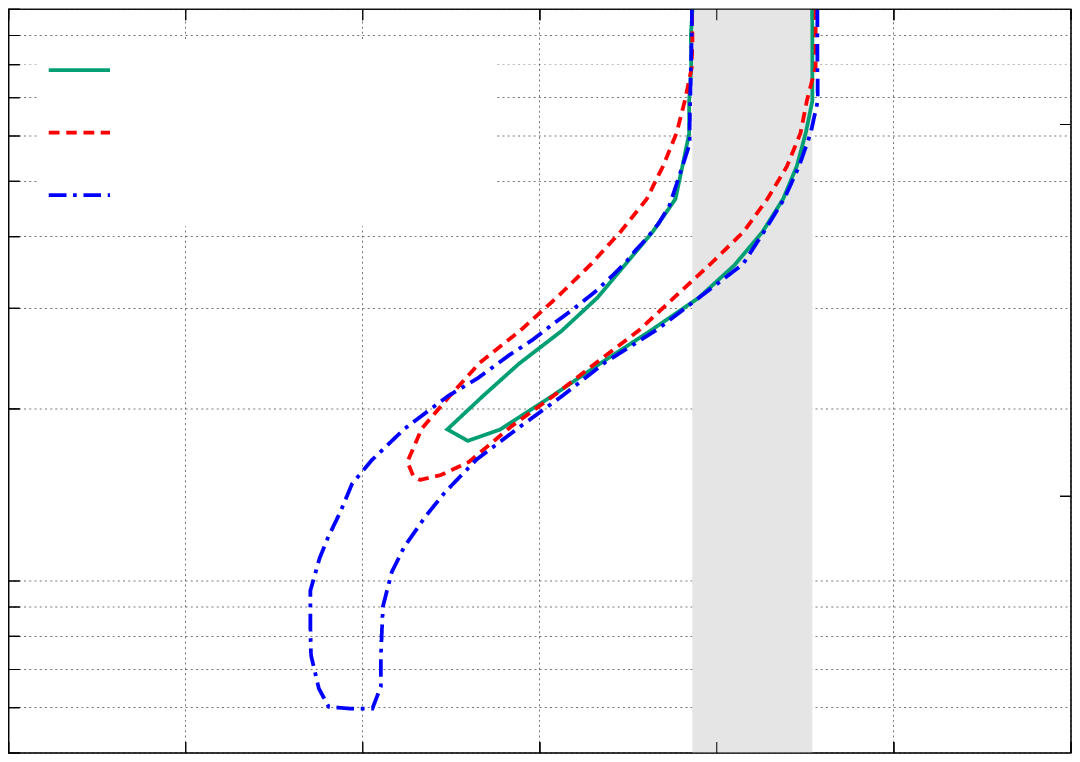}}
\vspace{0.5cm}
\end{center}
 \caption{
 \label{fig:osci_self_effect_radiative}
 Comparison of allowed regions in the ($\eta_B$, $T_{\rm RH}$) plane 
 in the cases with or without neutrino oscillation and self-interaction. The outside 
 of the small region surrounded by the contour is excluded at 95\%~C.L. in each case. 
 The constraint on $\eta_B$ at 95\% C.L. in the case of 
 the standard BBN is also shown as the gray-shaded region.
 We assume the 100\% radiative decays of $X$. The 95\% contour with the green solid 
 line is for the case with both neutrino oscillation and self-interaction, one with 
 the red long-dashed line is for the case only with neutrino oscillation, and one 
 with the short-dashed line is for the case only with neutrino self-interaction.
 }
\end{figure}

Figure~\ref{fig:allowed_region_radiative} shows the allowed region in the plane 
of $\eta_B$ and $T_{\rm RH}$ in the case of the $100\%$ radiative decays. 
In the current study, we assume that $\chi^2_{\rm D/H}$ and $\chi^2_{Y_p}$ 
follow Gaussian distribution. In this case, we can find the lower bound 
at $95$\%~C.L. on $T_{\rm RH}$ by requiring 
$\chi^2_{{\rm D/H}\, + \,Y_p}(\eta_B, T_{\rm RH}) < 5.991$:
\begin{eqnarray}
 T_{\rm RH} \gtrsim 1.8~{\rm MeV}\,,
\label{trmin_rad}
\end{eqnarray}
in the case with both neutrino oscillation and self-interaction. 
Also, we depict in Fig.~\ref{fig:osci_self_effect_radiative} 
the comparison between cases with and without neutrino oscillation 
or neutrino self-interaction. As can be seen from Fig.~\ref{fig:osci_self_effect_radiative}, 
we find $T_{\rm RH} \gtrsim 1.5~{\rm MeV}$ in the case with neutrino 
oscillation and without self-interaction, whereas $T_{\rm RH} 
\gtrsim 0.6~{\rm MeV}$ in the case without neutrino oscillation 
and with neutrino self-interaction. The BBN bound in the case 
with neutrino oscillation or self-interaction is tighter than 
that in the case without them. This is because, as we can see 
from Fig.~\ref{fig:trh_vs_D}, neutrino oscillation and 
self-interaction increase the value of $Y_p$, and the discrepancy 
between theoretical and observational values becomes large 
compared to the case without these effects.

\subsection{Results of BBN: Hadronic decay}
\begin{figure}[!t]
\vspace{1.5cm}
\begin{center}
\resizebox{16cm}{!}{\ \ \input{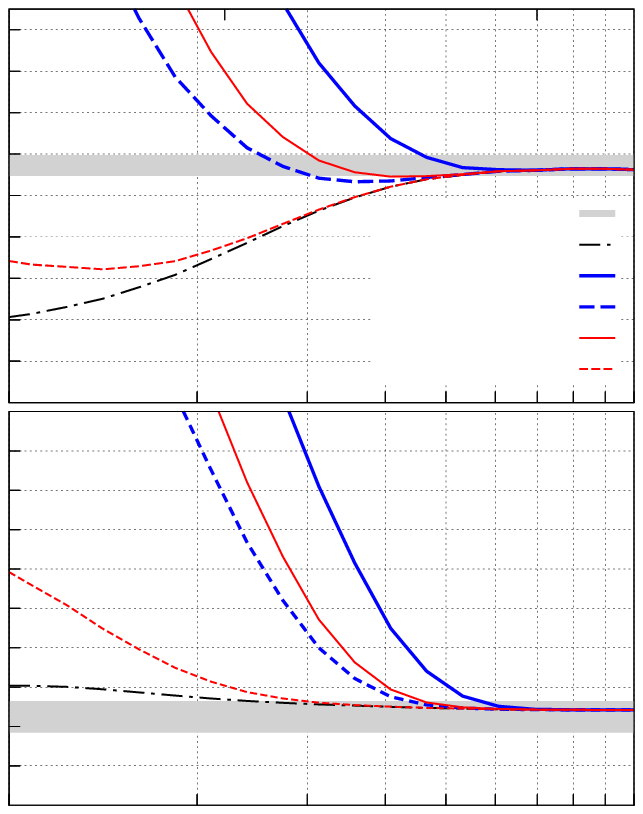}}
\end{center}
\vspace{0.5cm}
 \caption{
 \label{fig:trh_vs_D_and_Yp_w_had}
 D/H and $Y_p$ as a function of $T_{\rm RH}$ in the case with
 hadronic decays of $X$.  he blue thick solid- and dashed lines are
 for the case of $m_X = 10$~GeV, while the red thin solid- and
 dashed lines are for the case of $m_X = 100$~TeV. In addition, the
 blue- and red solid lines are for the case of Br $= 1$, while the
 blue- and red dashed lines are for the case of Br $= 0.001$. For
 comparison, we also plot the case of Br $= 0$ with the black
 dot-dashed line. In the figure, we consider both neutrino
 oscillation and self-interaction.
 }
\end{figure}

As described above, if the massive particles have a branching ratio
into hadrons, the constraint on $T_{\rm RH}$ imposed by BBN can be
modified compared to when the decays of $X$ are fully radiative ({\it i.e.} 
Br $= 0$).  The effects of hadronic decays on light element abundances
are shown in Fig.~\ref{fig:trh_vs_D_and_Yp_w_had} where we plot the
dependence of D and $^4$He abundances on $T_{\rm RH}$ for each value
of $m_X$ and Br. The case of Br $= 0$ in the figure corresponds to
the 100\% radiative decays of $X$, which is plotted for reference. In
the figure, we assume that the massive particles have a non-negligible
branching ratio into $u$\,$\bar{u}$ quark pairs to calculate the
number of hadrons produced in the decays of $X$ with Pythia~8.2
code.~\footnote{
  We have checked that the BBN bound does not depends on the 
  quark flavor emitted from the massive particles if the mass 
  of the massive particles is much larger than the total mass 
  of emitted quarks ({\it i.e.} $m_X >> m_{q_\alpha}$ where 
  $m_{q_\alpha}$ is the quark mass of particular flavor $\alpha$).
  }  

Figure~\ref{fig:trh_vs_D_and_Yp_w_had} shows that both 
the $^4$He and D abundances increases due to the hadronic decay 
effects. The reason is as follows. First, there exist more target 
protons than target neutrons in the system for $T \lesssim 
10$~MeV. This is because the neutron-proton ratio follows 
$n/p \simeq \exp(-Q/T)$ as long as the neutron-proton exchange 
reactions through weak interaction keep them in equilibrium, 
and $n/p$ is therefore smaller than unity in this epoch. 
Second, injected hadrons such as pions and kaons extraordinarily 
exchange ambient protons with neutrons through strong interaction 
via, {\it e.g.}
\begin{eqnarray}
  \label{eq:1}
  p + \pi^- &\to& n + \pi^0, \nonumber \\
  n + \pi^+ &\to& p + \pi^0. \nonumber
\end{eqnarray}
%
  We note that the neutral pion $\pi^0$ produced in the processes 
  immediately decays into two photons and does not cause the corresponding 
  inverse processes in this epoch.
%
%
For these reasons, the injected hadrons induce a net flow from $p$ to $n$. 
This gives an out of equilibrium abundances of neutron and proton and leads 
to a larger $n/p$ ratio (see \cite{Reno_1988,Kawasaki_2000} for 
more detailed discussions). As a result, the $^4$He and D abundances, 
which increase with $(n/p)_{\rm f}$, get larger than those of 
the standard BBN. In addition, it can be seen from 
Fig.~\ref{fig:trh_vs_D_and_Yp_w_had} that the effect of
hadronic decays is large for a large Br or a small $m_X$. Comparing the
cases of $m_X = 100$~TeV, Br = 0.001 (red thin dashed) and
$m_X = 10$~GeV, Br = 1 (blue thick solid) in Fig.~\ref{fig:trh_vs_D_and_Yp_w_had}, 
we can see that the discrepancy of D/H or $Y_p$ between these cases 
are of the same order or much larger than that of the $2\sigma$ 
observational error if $T_{\rm RH}$ is a few MeV.

To understand the reason, we define the comoving variable for the
initial abundance of the massive particles $Y_X = n_X/s$ where $n_X$
is the number density of the massive particles $X$, and $s$ is the
total entropy density of the Universe.  If we assume that $X$
dominates the total energy at the initial time and most of it is
transferred to radiation components before the reheating is
completed, we can write the initial value of $Y_X$ as follows:
\begin{equation}
 Y_X = \frac{n_X}{s} \sim 
  \frac{(\frac{\pi^2}{30}\, g^*\,T_{\rm RH}^4)/m_X}{\frac{2\pi^2}{45}\, g^*_s\, 
  T_{\rm RH}^3} \sim \frac{3}{4} \frac{T_{\rm RH}}{m_X} \,,
\end{equation}
where $g^*$ and $g^*_s$ are relativistic degrees of freedom defined by
energy and entropy density respectively. In the standard big-bang
cosmology, $g^* \sim g^*_s$ holds before electron-positron
annihilation sets in.  As we can see from the above expression, $Y_X$
gets larger for smaller $m_X$.  In addition, the number of hadrons
emitted from the decays of $X$ is almost proportional to $m_X^{0.4}$
(see Ref.~\cite{Kawasaki_2005}), and therefore the total number of
hadrons emitted from $X$ is almost proportional to $m_X^{-0.6}$.
Since the energetic hadrons instantaneously lose their energy and are
thermalized with background particles before inter-converting ambient
neutrons and protons, the number of emitted hadrons determine the
magnitude of the hadronic-decay effect on BBN. Therefore, the
influence of hadronic decays on BBN should be stronger for smaller
$m_X$.

\begin{figure}[!t]
\vspace{0.5cm}
\begin{center}
\resizebox{14cm}{!}{\input{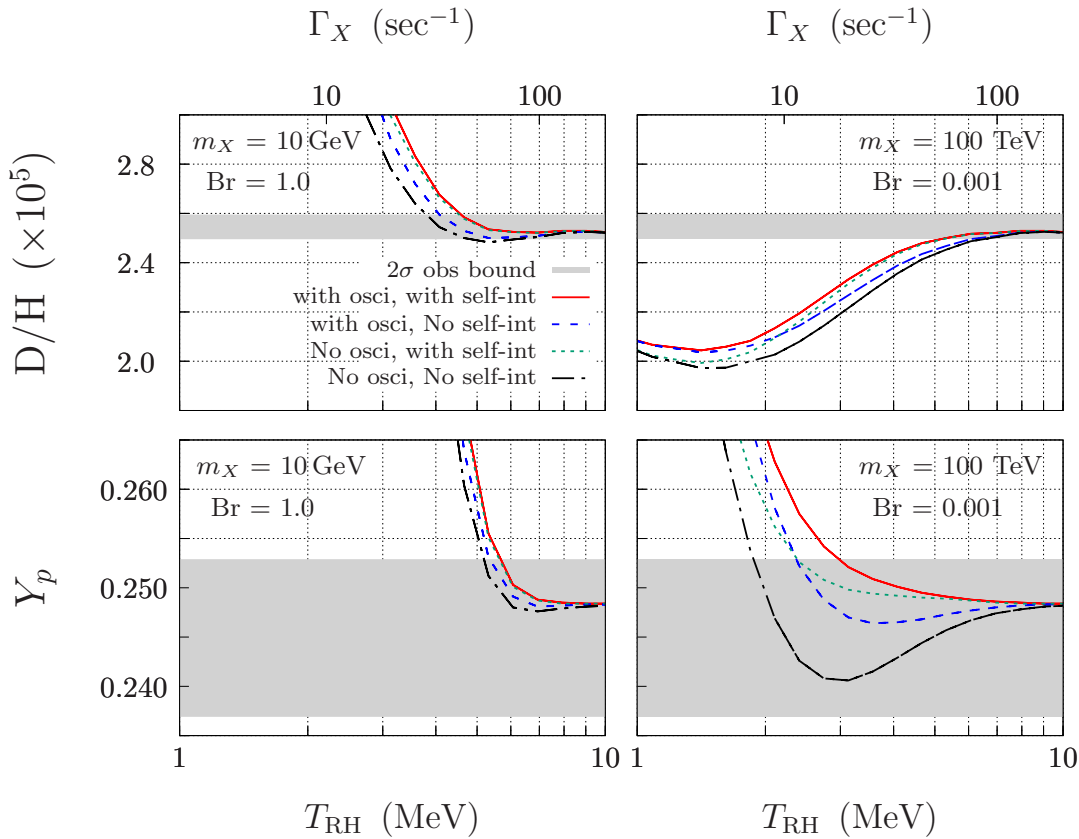}}
\end{center}
\vspace{0.5cm}
 \caption{
 \label{fig:osci_self_effect_had_D}
 Effects of neutrino oscillation and self-interaction on D/H and $Y_p$ 
 in the case with hadronic decays of $X$. We adopt $\eta_B = 6.13 \times 10^{-10}$ 
 in the figure. The red solid- and blue long-dashed lines are for the case with neutrino 
 oscillation, while the green short-dashed and black dot-dashed line are for the case 
 with neutrino self-interaction.
 }
\end{figure}

For the purpose of showing the effects of neutrino oscillation and self-interaction 
on the light element abundances in the case of hadronic decays, we plot in Fig.~11 
the dependence of $Y_p$ and D/H on $T_{\rm RH}$ for ($m_X$, Br) = (10~GeV, 1.0) 
and (100~TeV, 0.001) where we expect large and small effects of hadronic decays, 
respectively. As we can see from Fig.~11, 
if $T_{\rm RH}$ is a few MeV, neutrino oscillation and self-interaction affect 
light element abundances at the level of ${\cal O}(10)$\% for D/H and ${\cal O}(1)$\% for $Y_p$ 
when $m_X = 10$~GeV and ${\rm Br} = 1.0$, whereas the correction is 
${\cal O}(10)$\% for both cases of D and $Y_p$ when $m_X = 100$~TeV 
and ${\rm Br} = 0.001$. Since we give the observational bound on 
$T_{\rm RH}$ by summing up the $\chi^2$ values of D/H and $Y_p$, 
the constraint on $T_{\rm RH}$ should be changed by 
${\cal O}(1)$\% when $m_X = 10$~GeV and ${\rm Br} = 1.0$ and by 
${\cal O}(10)$\% when $m_X = 100$~TeV and ${\rm Br} = 0.001$. 

\begin{figure}[!t]
\begin{center}
\resizebox{13cm}{!}{\!\!\!\!\!\!\!\!\!\!\!\!\!\!\!\!\!\!\input{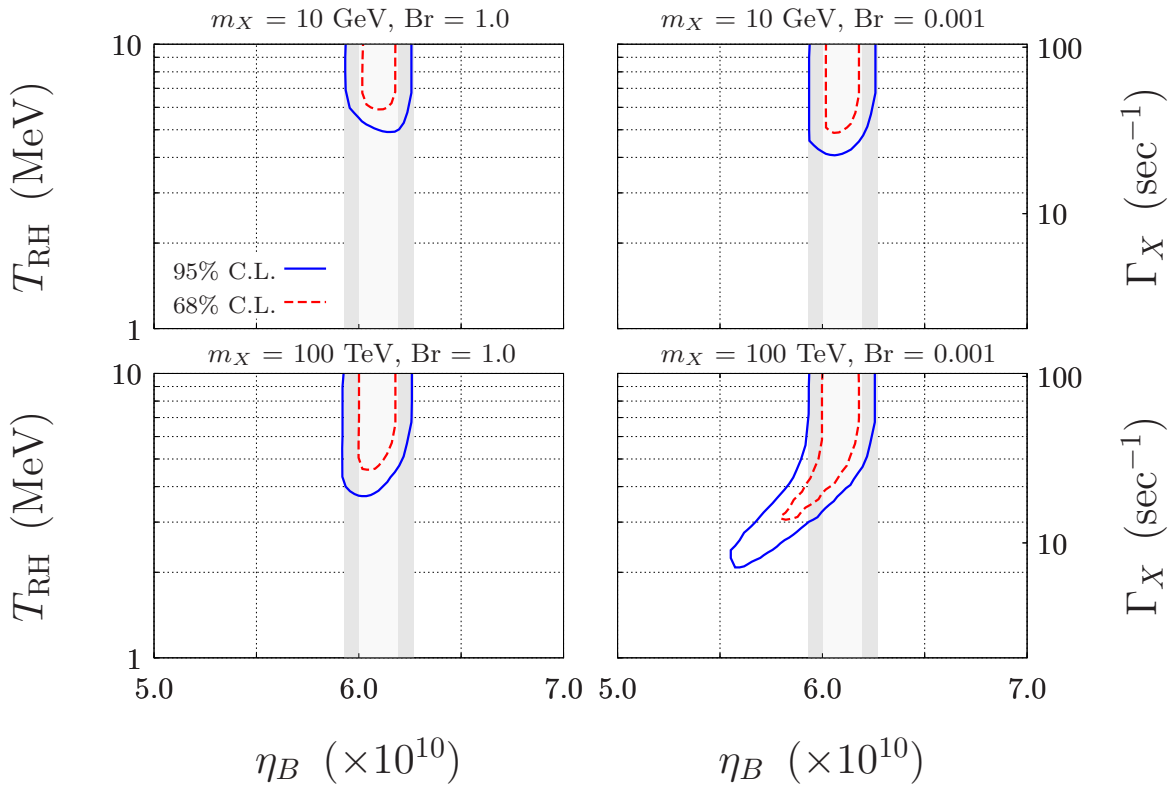}}
\end{center}
 \caption{
 Same as Fig.~\ref{fig:allowed_region_radiative}, but for the case of hadronic decays of 
 the massive particles. The outside of the small region surrounded by the blue solid 
 (red dashed) line is excluded at 95\%~(68\%)~C.L. The constraint on $\eta_B$ at 95\% (68\%) 
 C.L. in of hadronic decaythe case of the standard BBN is also shown as the dark (light) shaded region. 
 }\label{fig:allowed_region_hadronic}
\end{figure}

We show in Fig.~12 
the allowed region in the same plane as Fig.~\ref{fig:allowed_region_radiative}, 
but in the case when hadronic decays are included. In the figure, 
we show four representative cases of ($m_X$, Br) = (10~GeV, 1.0), 
(10~GeV, 0.001), (100~TeV, 1.0) and (100~TeV, 0.001). 

A possible minimum value of the reheating temperature $T_{\rm RH,\,min}$ 
in terms of BBN is shown in Fig.~13 
as a function of $m_X$. We can see from the figure that the BBN bound
is tighter in the case of a small $m_X$ or a large Br. Consequently,
we obtain the lower bound on $T_{\rm RH}$ at $95$\%~C.L.: 
\begin{equation}
 T_{\rm RH}\, \gtrsim\, 4.1-4.9~{\rm MeV}\ \ \ {\rm for}\ \ m_X = 10~{\rm GeV}-100~{\rm TeV}\,,  
\end{equation}
when the hadronic branching ratio Br $= 1.0$, whereas 
\begin{equation}
 T_{\rm RH}\, \gtrsim\, 2.1-3.7~{\rm MeV}\ \ \ {\rm for}\ \ m_X = 10~{\rm GeV}-100~{\rm TeV}\,,
\end{equation}
when Br $= 0.001$ in the case with both neutrino oscillation 
and neutrino self-interaction. In addition, we find neutrino 
oscillation and neutrino self-interaction can change the value 
of $T_{\rm RH,\, min}$ at the level of ${\cal O}(1)$\% for most 
of the range of $m_X$ in the case of hadronic decays. 

\begin{figure}[!t]
\begin{center}
\resizebox{16cm}{!}{\ \ \ \ \ \ \ \ \ \ \ \input{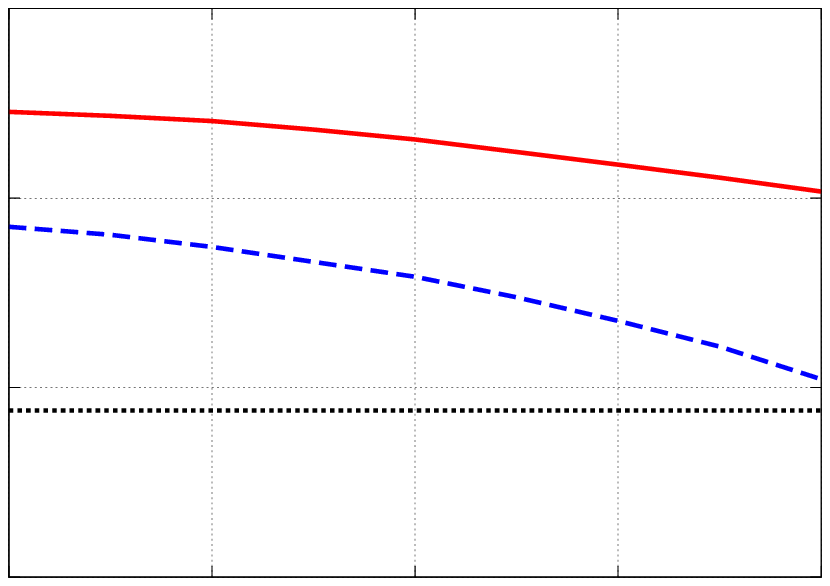}}
\end{center}
\vspace{0.5cm}
 \caption{
 Lower bound on the reheating temperature $T_{\rm RH,\,min}$ at $95\%$~C.L. 
 as a function of $m_X$ in the case with both neutrino oscillation and neutrino self-interaction. 
 We plot the results in the cases of Br $= 0.001$ (blue dashed) and $1.0$ (red solid). 
 The 100\% radiative decay case is also plotted with the black dashed line.
 }\label{fig:mx_vs_trhmin}
\end{figure}
\section{\label{Sec:conclusion}Conclusion}
In this paper, we have investigated the possibility that the reheating
temperature of the Universe is ${\cal O}$(1)~MeV motivated by long-lived 
massive particles which often appear in the particle physics theory beyond 
the standard model and induce a late-time entropy production by their decays. 
In this scenario, neutrinos are not necessarily thermalized well before the 
beginning of BBN. Hence, the expansion rate of the Universe and weak reaction 
processes are significantly altered, which changes the freezeout value of 
the neutron to proton ratio. We have calculated the thermalization process
of neutrinos including effects of both neutrino oscillation and
neutrino self-interaction (Figs.~\ref{fig:trh_vs_neff}--\ref{fig:neff_evolution2}), 
and obtained a lower bound on the reheating temperature 
$T_{\rm RH} \gtrsim 1.8$~MeV (95\%~C.L.) (Fig.~\ref{fig:allowed_region_radiative}) 
in the case of the 100$\%$ radiative decay.

On the other hand, if the massive particles also decay into hadrons, 
there is an additional effect on BBN via inter-conversion of ambient 
neutron and proton through the scatterings of the hadrons. In this case, 
the constraint becomes tighter than that of the 100$\%$ radiative decay 
(Fig.~\ref{fig:trh_vs_D_and_Yp_w_had}). Then, we obtained the lower bound 
$T_{\rm RH} \gtrsim$ 2~MeV--5~MeV (95\%~C.L.) depending on the masses 
of the massive particles (10~GeV--100~TeV) and the hadronic branching 
ratio of the decay (Figs.~12--13). 

In addition, we found that neutrino oscillation and neutrino
self-interaction increase the efficiency of neutrino thermalization
(Figs.~\ref{fig:trh_vs_neff} and \ref{fig:neff_evolution2}) and
decrease the exchange rate between neutrons and protons, thereby
enhancing the theoretically expected abundances of helium, $Y_p$
(Fig.~\ref{fig:trh_vs_Yp}), and deuterium, D/H (Fig.~\ref{fig:trh_vs_D}).  
These effects increase the minimum value of the reheating temperature 
at the level of ${\cal O}(10)$\% in the case of the 100\% radiative decays 
(Fig.~\ref{fig:allowed_region_radiative}) and ${\cal O}(1)$\% in most 
cases of hadronic decay of hadronic decays (Fig.~11). 

Finally, let us comment on the future prospects of this study. 
This time, we only focused on BBN to constrain $T_{\rm RH}$. 
On the other hand, as described in Sec.~\ref{Sec:intro}, 
CMB and LSS also depend on the expansion rate in the Universe, 
and therefore have a sensitivity to the neutrino thermalization. 
In addition, the recombination history depends on the $^4$He 
abundance which is strongly affected by the hadronic decay effects.
Therefore, theoretical results of CMB and LSS should be different 
from those in the case of radiative decay. 
We will discuss observational constraints on $T_{\rm RH}$ 
from CMB and LSS in addition to BBN assuming hadronic decays of 
the massive particles in a forthcoming paper~\cite{Hasegawa_2019}.

\begin{acknowledgments}
 Numerical computations were carried out on PC clusters at Center 
 for Computational Astrophysics, National Astronomical Observatory of Japan (NAOJ) 
 and Computing Research Center, High Energy Accelerator Research Organization (KEK).
 KK is supported by JSPS KAKENHI Grants No.~JP17H01131, MEXT Grant-in-Aid 
 for Scientific Research on Innovative Areas Nos.~JP15H05889, JP18H04594, JP19H05114, 
 and by WPI, MEXT, Japan. The work of RSLH, TT, and SH is supported by the Villum Foundation.
\end{acknowledgments}
\bibliography{ref}
\bibliographystyle{jhep.bst}
%

\end{document}